\documentclass[aps,pra,11pt,superscriptaddress,nofootinbib]{revtex4-2}
\newcommand{\figurewidth}{0.5\linewidth}

\usepackage{amsmath,amssymb}
\usepackage{graphicx,enumitem}
\usepackage[normalem]{ulem}
\usepackage[dvipsnames]{xcolor}
\usepackage[colorlinks,allcolors=Blue,linktocpage]{hyperref}
\usepackage[boxsize=.5em, aligntableaux=center]{ytableau}
\usepackage{longtable,placeins}
\usepackage{newtxtext,newtxmath,calrsfs}
\usepackage{CJK}
\setlist{align=parleft,leftmargin=0pt,itemindent=2.5\parindent,labelsep=0em,labelwidth=1.5\parindent,nosep}

\newcommand{\rd}{\mathrm{d}}
\newcommand{\rA}{\mathrm{A}}
\newcommand{\rU}{\mathrm{U}}
\newcommand{\rT}{\mathrm{T}}
\newcommand{\br}{\mathbf{r}}
\newcommand{\bbI}{\mathbb{I}}
\newcommand{\Sp}{\mathrm{Sp}}
\newcommand{\SO}{\mathrm{SO}}
\newcommand{\SU}{\mathrm{SU}}
\newcommand{\bY}{\bar{Y}}
\newcommand{\bDelta}{\boldsymbol{\Delta}}
\newcommand{\bPsi}{\boldsymbol{\Psi}}
\newcommand{\bOmega}{\boldsymbol{\Omega}}
\newcommand{\sm}{\hphantom{-{}}}
\newcommand{\Norb}{N_\mathrm{orb}}

\begin{document}

\begin{CJK*}{UTF8}{bkai}

\title{\boldmath A new series of 3D CFTs with $\Sp(N)$ global symmetry on fuzzy sphere}
\author{Zheng Zhou (周正)}
\affiliation{Perimeter Institute for Theoretical Physics, Waterloo, Ontario N2L 2Y5, Canada}
\affiliation{Department of Physics and Astronomy, University of Waterloo, Waterloo, Ontario N2L 3G1, Canada}
\author{Yin-Chen He}
\affiliation{Perimeter Institute for Theoretical Physics, Waterloo, Ontario N2L 2Y5, Canada}

\begin{abstract} 
    The quest to discover new 3D CFTs has been intriguing for physicists. For this purpose, fuzzy sphere regularisation that studies interacting quantum systems defined on the lowest Landau level on a sphere has emerged as a powerful tool. In this paper, we discover a series of new CFTs with global symmetry $\Sp(N)$ in the fuzzy sphere models that are closely related to the $\SO(5)$ deconfined phase transition, and are related to a $\Sp(N)/(\Sp(M)\times \Sp(N-M))$ non-linear sigma model with a Wess-Zumino-Witten term. We numerically verify the emergent conformal symmetry by observing the integer-spaced conformal multiplets, the quality of conformal generators, and studying the finite-size scaling of the conformality. We discuss possible candidates for these newly discovered CFTs, the most plausible ones being Chern-Simons-matter theories which have $N$ flavours of gapless bosons or fermions coupled to a non-Abelian (\textit{viz.} $\Sp(1)$, $\Sp(2)$, \textit{etc.}) Chern-Simons gauge field. Our work provides new avenues for studying interacting CFTs in 3D, possibly facilitating the non-perturbative study of critical gauge theories and previously undiscovered CFTs. 
\end{abstract}

\maketitle

\end{CJK*}

\textit{Introduction.} ---
Conformal field theory (CFT) is one of the central topics of modern physics. It has provided invaluable insights into critical phenomena in condensed matter physics~\cite{Polyakov1970Conformal,Cardy1996Scaling}, string theory and AdS/CFT correspondence in quantum gravity~\cite{Maldacena1998AdSCFT}, and enhanced our understanding of the renormalisation group~\cite{Zamolodchikov1986Irreversibility} and other fundamental structures and dynamics of quantum field theory (QFT). In 2d, many CFTs are well understood thanks to their integratibility~\cite{DiFrancesco1997CFT,Belavin1984BPZ}. Going to higher dimensions, CFTs are much less well-studied due to a much smaller conformal group. Existing approaches, including Monte Carlo lattice simulation and numerical conformal bootstrap~\cite{Poland2019Bootstrap,Rychkov2023Bootstrap}, despite having achieved many successes, can only handle a limited number of CFTs and obtain a limited number of conformal data.

Therefore, the quest to discover new CFTs in $d\ge 3$ has been particularly intriguing to physicists. A virgin land on this quest is the parity-breaking CFTs. In 3D, the Chern-Simons-matter theory stands out as the most well known and possibly the only known type of parity-breaking CFTs. These theories involve gapless bosonic or fermionic fields coupled to a Chern-Simons gauge field. The Chern-Simons-matter theories play a crucial role in understanding phase transitions of topological orders~\cite{wen1993transitions,Chen1993,Yang2008Feshbach,Maissam2014_FQHtransition,QED3CS_Lee,ZouQCD3,Song_MoireTransition}. They are also conjectured to exhibit interesting field theory dualities~\cite{Seiberg2016Duality,Karsh2016Duality,Hsin2016LevelRank,Aharony2016Duality,Benini2017Duality}. Despite their fundamental and practical importance, there has been no non-perturbative study for these parity-breaking theories, primarily due to the intrinsic sign problem they pose for Monte Carlo simulations.

Recently, fuzzy sphere regularisation has emerged as a new powerful method to study 3D CFTs~\cite{Zhu2023Uncovering}. By studying interacting quantum systems on the fuzzy (non-commutative) sphere~\cite{Madore1992Fuzzy}, the method realises $(2+1)$D quantum phase transitions on the geometry $S^2\times\mathbb{R}$. Compared with conventional methods that involve simulating lattice models, this approach offers distinct advantages including exact preservation of rotation symmetry, direct observation of emergent conformal symmetry, and the efficient extraction of conformal data. In the fuzzy sphere method, the state-operator correspondence~\cite{Cardy1984Conformal,Cardy1985}  plays an essential role. Specifically, there is a one-to-one correspondence between the eigenstates of the critical Hamiltonian on the sphere and the CFT operators, where the energy gaps are proportional to the scaling dimensions. The power of this approach has been demonstrated in the context of the 3D Ising transition~\cite{Zhu2023Uncovering,Hu2023Operator,Han2023FourPoint,Hu2024FFunction,Hofmann2023MonteCarlo}, where the presence of emergent conformal symmetry has been convincingly established and a wealth of conformal data has been accurately computed. The study has been extended to the magnetic line defect~\cite{Hu2023Defect,Zhou2024Overlap,Cuomo2024Cusp}, various conformal boundaries in 3D Ising CFT~\cite{Zhou2024Boundary,Dedushenko2024Boundary}, conformal generators~\cite{Fardelli2024Generator,Fan2024Generator}, Wilson-Fisher theory~\cite{Han2023O3}, and $\SO(5)$ deconfined criticality~\cite{Zhou2024SO5}. 

Among these researches, the $\SO(5)$ deconfined quantum critical point (DQCP)~\cite{Senthil2004DQCP,Senthil2023DQCP} is in particular noticeable. This pioneering example of phase transitions beyond the Landau paradigm can be described by the non-linear sigma model (NL$\sigma$M) on the Grassmannian $\Sp(2)/(\Sp(1)\times\Sp(1))\equiv S^4$ with a level-1 Wess-Zumino-Witten (WZW) term~\cite{Nahum2015SO5}. This theory has several dual descriptions, \textit{viz.} $\SU(2)$ QCD with two flavours of fermions, and $\mathbb{C}\mathrm{P}^1$ theory, \textit{i.e.}, scalar QED with two flavours of complex boson~\cite{Senthil2004DQCP,Motrunich2004CP1,Senthil2004Quantum,Wang2017DQCP}. It can be realised on the fuzzy sphere by half-filling four flavours of fermions and restricting the symmetry to $\SO(5)\equiv\Sp(2)/\mathbb{Z}_2$~\cite{Zhou2024SO5,Ippoliti2018TorusSO5,Wang2021TorusSO5,Chen2024SO51,Chen2024SO52}. As the parity symmetry is realised on the fuzzy sphere as the particle-hole symmetry, a generalisation of this model that breaks the particle-hole symmetry is a promising candidate for realising parity-breaking CFTs. Specifically, we consider $2N$ flavours of fermions, fill $2M$ of these flavours, and break the symmetry from $\SU(2N)$ to $\Sp(N)$~\footnote{Here we adopt a notation that $\Sp(N)$ denotes the group of $2N\times2N$ unitary symplectic matrices.}. This model should be described by a NL$\sigma$M on Grassmannian
\begin{equation*}
    \frac{\Sp(N)}{\Sp(M)\times\Sp(N-M)}
\end{equation*}
with a level-1 WZW term. This WZW-NL$\sigma$M is known to be closely related to Chern-Simons-matter theories, particularly when $N\neq 2M$~\cite{Komargodski2018QCD}. Although the NL$\sigma$M itself is non-renormalisable, it can be used as a tool to help determine the phase diagram. By matching the global symmetry and the anomaly, one can figure out the natural candidate IR fates, including CFTs and TQFTs, that the fuzzy sphere model could flow to. Different values of $N$ and $M$ correspond to different unexplored CFTs.

In this paper, we study these partially filled $\Sp(N)$ models with $(N,M)=(3,1),(4,1)$ and half-filled theories $(N,M)=(2,1),(4,2)$ numerically. By tuning the interaction strength, we find a parameter region for each $(N,M)$ conformal symmetry, with evidence such as conserved currents, integer-spaced conformal multiplets, and conformal generators with good quality. By defining a cost function, we further find hint that the spectrum scales towards conformality with currently available system sizes for the $\Sp(3)$ and $\Sp(4)$ models, which is distinct from the case of $\Sp(2)$ --- the likely-pseudocritical $\SO(5)$ DQCP~\cite{Wang2017DQCP,Gorbenko2018Walking,Gorbenko2018Walking2}. These evidences show that each of the $(N,M)$ fuzzy sphere models at the particular parameters is described by a CFT. We extract and analyse one of the most important property of the CFT --- the scaling dimensions of primary operators --- through state-operator correspondence. Different values of $(N,M)$ flows to different CFTs, and these CFTs are likely to be unexplored previously. By matching the anomaly through the corresponding NL$\sigma$M, we show that the most plausible candidates are the Chern-Simons-matter theories which have $N$ flavours of gapless bosons or fermions coupled to a non-Abelian (\textit{viz.} $\Sp(1)$, $\Sp(2)$, \textit{etc.}) Chern-Simons gauge field~\cite{Komargodski2018QCD}. Our work provides new avenues for studying interacting CFTs in 3D, possibly facilitating the non-perturbative study of critical gauge theories and previously undiscovered CFTs.

\textit{Model.} ---
We consider $2N$ flavours of fermions $\psi_a(\mathbf{r})(a=1,\dots,2N)$ moving on a sphere in the presence of a $4\pi s$-monopole at its centre. Due to the presence of the monopole, the fermion single-particle eigenstates form highly degenerate quantised Landau levels. The ground state, \textit{i.e.}, lowest Landau level (LLL), has a degeneracy $\Norb=2s+1$ for each flavour. We partially fill the LLL and set the gap between the LLL and higher Landau levels to be much larger than other energy scales in the system. In this case, we can effectively project the system into the LLL. After the projection, the fermion operator $\psi_a(\mathbf{r})(a=1,\dots,2N)$ can be expressed in terms of the creation and annihilation operators on the LLL $\psi_a(\mathbf{r})=\sum_{m=-s}^sY_{sm}^{(s)}c_{ma}$.  Furthermore, we fill $2M$ out of the $2N$ flavours to obtain the desired NL$\sigma$M.

To construct an interaction Hamiltonian that breaks the maximal flavour symmetry of $\SU(2N)$ down to $\Sp(N)$, we consider the $\Sp(N)$-invariant fermion bilinears: $\psi^\dagger_a\psi_a$ is $\SU(2N)$-invariant; $\Omega_{aa'}\psi_a\psi_{a'}$ where
\begin{equation*}
    \Omega=\begin{pmatrix}0&\bbI_N\\-\bbI_N&0\end{pmatrix}
\end{equation*}
is $\Sp(N)$ invariant but not $\SU(2N)$-invariant. Consequently, we construct
\begin{multline}
    H=\int\rd^2\br_1\,\rd^2\br_2\,\left(U(\br_{12})\psi^\dagger_a(\br_1)\psi^\dagger_b(\br_2)\psi_b(\br_2)\psi_a(\br_1)\right.\\\left.-\frac{1}{2} V(\br_{12})\Omega_{aa'}\Omega_{bb'}\psi^\dagger_a(\br_1)\psi^\dagger_{a'}(\br_2)\psi_{b'}(\br_2)\psi_b(\br_1)\right).
    \label{eq:hmt}
\end{multline}
The potential functions $U(\br_{12})$ and $V(\br_{12})$ can be parametrised by the Haldane pseudopotentials~\cite{Haldane1983Pseudopotential} $U_l$ and $V_l(l=0,\dots,2s)$. Only the $V_l$ with even-$l$ keeps the $\Sp(N)$ symmetry. This model is a direct generalisation of the $\SO(5)$ DQCP on the fuzzy sphere or Landau levels~\cite{Zhou2024SO5,Ippoliti2018TorusSO5,Wang2021TorusSO5,Chen2024SO51,Chen2024SO52}, and for detailed construction of the invariant fermion bilinears, the pseudopotentials and the projection onto LLL, see Appendix~\ref{app:model}. In the calculations below, we set $U_0=1$ as the energy unit and vary $U_1,U_2,V_0,V_2$ as the tuning parameters. 

\begin{figure}[b]
    \centering
    \includegraphics[width=\figurewidth]{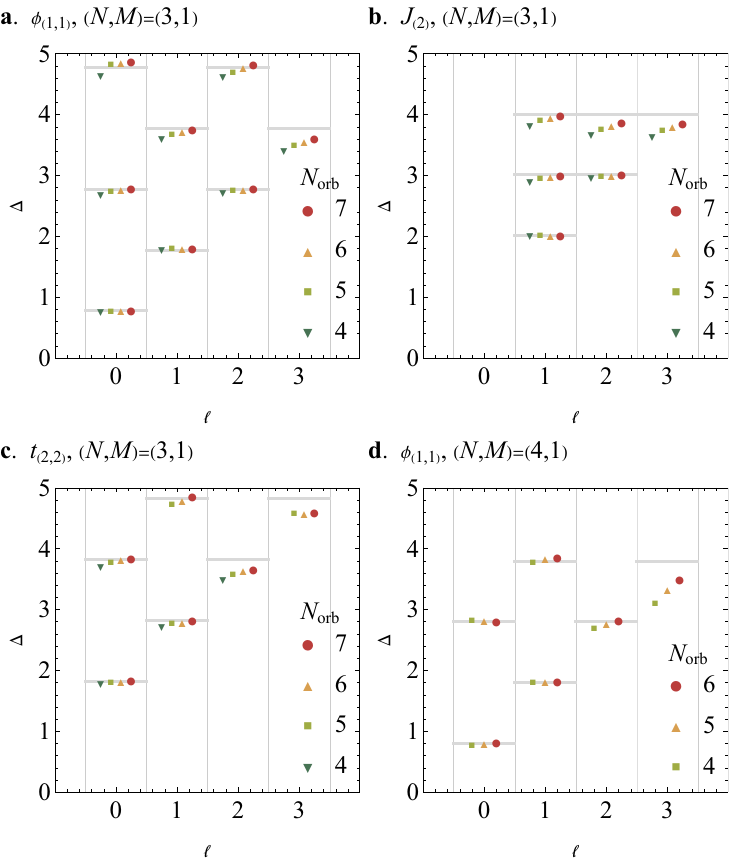}
    \caption{The scaling dimensions of the conformal multiplet of (a) $\phi_{(1,1)}$, (b) $J^\mu_{(2)}$, (c) $t_{(2,2)}$ in the $(N,M)=(3,1)$ theory and (d) $\phi_{(1,1)}$ in the $(N,M)=(4,1)$ theory at different spin $\ell$ and system size $\Norb$. The horizontal grey bars denote the anticipated values based on the integer-spaced level from the conformal symmetry. The optimal points at the maximal sizes are taken for the calculation. }
    \label{fig:multip}
\end{figure}

\textit{Results.} ---
We perform an exact diagonalisation (ED) calculation for the lowest 800 eigenstates with a maximal size $\Norb=7,6,9,5$ for $(N,M)=(3,1),(4,1),(2,1),(4,2)$ respectively using our open source Julia package FuzzifiED~\cite{FuzzifiED}. As stated by the state-operator correspondence~\cite{Cardy1984Conformal,Cardy1985}, for a CFT, each eigenstate corresponds to a scaling operator. Its scaling dimension is proportional to the excited energy $\Delta_i=\alpha(E_i-E_0)$ where the undetermined coefficient $\alpha$ depends on the microscopic model. Its representation under $\SO(3)$ rotation symmetry can be measured by the expectation value of the total angular momentum $\langle\hat{L}^2\rangle=\ell(\ell+1)$, and its representation under $\Sp(N)$ global symmetry can be measured by the expectation value of the quadratic Casimir $C_2$~\cite{Yamatsu2015Lie}. For detailed construction, see Appendix~\ref{app:l2c2}. 

\begin{figure}[b]
    \centering
    \includegraphics[width=\figurewidth]{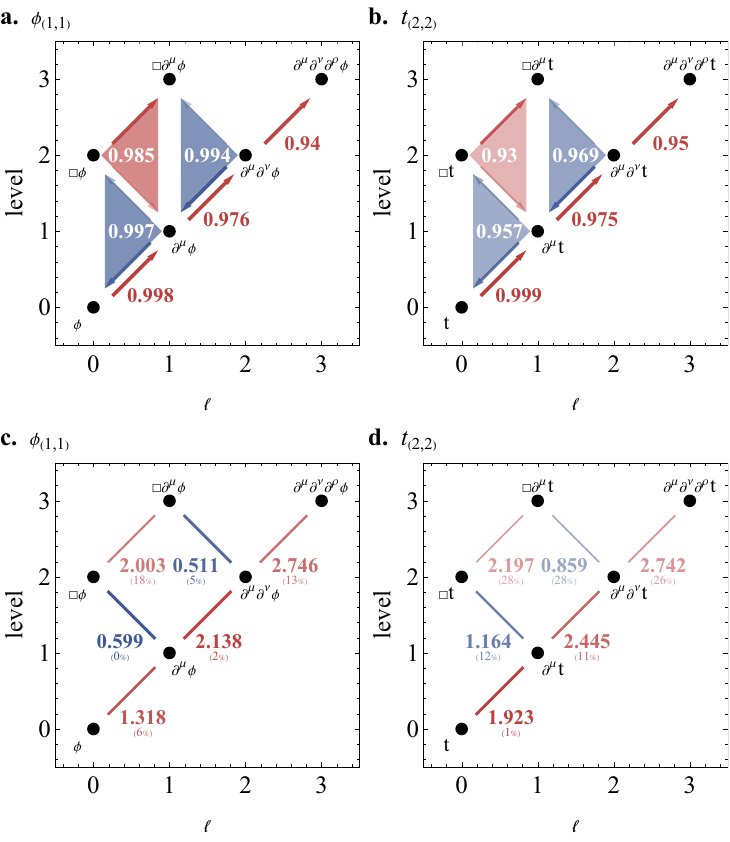}
    \caption{(a,b) The total squared overlap of the normalised state $(\Lambda^\mu|\partial^n\Phi,\ell\rangle)_{\ell'}$ with $|\partial^{n+1}\Phi,\ell'\rangle$ and $|\partial^{n-1}\Phi,\ell'\rangle$ for $\ell'=\ell\pm 1$ for the multiplets of (a) $\Phi=\phi_{(1,1)}$ and (b) $\Phi=t_{(2,2)}$. Numbers closer to unity represent better conformal symmetry. The tail of the arrow represents the state that $\Lambda^\mu$ acts upon and the head represents the state with which the overlap is taken. (c,d) The matrix elements $\langle\partial^{n'}\Phi,\ell'|\Lambda^z|\partial^n\Phi,\ell\rangle$ and their deviation from conformality. The opacity and thickness represent the closeness to conformality. The colours are for better readability. The calculations are done for $(N,M)=(3,1)$ theory and $\Norb=6$.}
    \label{fig:gen}
\end{figure}

A spectrum of CFT is characterised by (1) conserved symmetry currents $J^\mu$ and stress tensor $T^{\mu\nu}$ with scaling dimensions $\Delta_{J^\mu}=2$ and $\Delta_{T^{\mu\nu}}=3$; (2) conformal descendants with integer spacing from the primary operator. To analyse an energy spectrum described by a CFT, we follow the recipe below~\cite{Zhu2023Uncovering}: (1) we select the lowest state in each representation and identify it as a primary; (2) we identify its descendants by matching the quantum numbers $\ell,C_2$ and $\Delta$; (3) we then remove the identified conformal multiplet from the spectrum and repeat the process. To search for a parameter point with conformal symmetry, we define a cost function $Q$ as the root mean square of the deviations of the 2 conserved currents and 6 conformal descendants with $\Delta<3$ from the prediction of conformal symmetry. We minimise the cost function with respect to the parameters $U_1,U_2,V_0,V_2$ in the Hamiltonian and the coefficient $\alpha$ to find the ideal point for the calculation. We take the optimal point at the maximal size for future calculation. For details, see Appendix~\ref{app:cost}. 

The good conformal symmetry is evidenced by the existence of integer-spaced conformal multiplets. For a scalar primary $\Phi$, its descendants are of the form $\partial^\ell\Box^n\Phi$ with spin-$\ell$ and $\Delta=\Delta_\Phi+n+2l$; for the conserved current $J^\mu$, it has two forms of descendants $\Box^n\partial^{\mu_1}\dots\partial^{\mu_{\ell-1}}J^{\mu_\ell}$ and $\Box^n\partial^{\mu_1}\dots\partial^{\mu_{\ell-1}}\partial^\nu\epsilon_{\mu_\ell\nu\rho} J^\rho$~\cite{Zhu2023Uncovering}. All the operators in the spectrum can be organised into multiplets. We take the lowest three primaries as an example, \textit{viz.} the lowest operator $\phi_{(1,1)}$ in the $\ydiagram{1,1}$ representation of $\Sp(N)$, the lowest operator $t_{(2,2)}$ in the $\ydiagram{2,2}$ representation of $\Sp(N)$, and the conserved symmetry current $J_{(2)}^\mu$ where the subscripts describe the Young diagrams for the representations, and these primaries will be explained in detail later. We identify all of their lowest-lying descendants with $\Delta<5$ and $\ell\leq 3$ in the $(N,M)=(3,1)$ and $(4,1)$ theories, and the scaling dimensions are very close to the integer spacing with the primary operator~(Fig.~\ref{fig:multip}).

Further evidence for conformal symmetry comes from the conformal generators on the fuzzy sphere~\cite{Fan2024Generator,Fardelli2024Generator}. The conformal generator $\Lambda^\mu=P^\mu+K^\mu$ on the states is the Hamiltonian density $\mathcal{H}(\br)$ integrated against $l=1$ spherical harmonics $Y_{1m}$. The detailed construction can be found in Appendix~\ref{app:gen}. For a state $|\partial^n\Phi,\ell\rangle$ in the multiplet of primary $\Phi$, the action of $\Lambda^z$ brings it to the superposition of $|\partial^{n\pm 1}\Phi, \ell'=\ell\pm 1\rangle$, we decompose $\Lambda^z|\partial^n\Phi,\ell\rangle$ into angular momentum sectors and check the overlap 
\begin{equation}
    \frac{\left|\langle\partial^{n+1}\Phi, \ell'|\left(\Lambda_z|\partial^n\Phi,\ell\rangle\right)_{\ell'}\right|^2+\left|\langle\partial^{n-1}\Phi, \ell'|\left(\Lambda_z|\partial^n\Phi,\ell\rangle\right)_{\ell'}\right|^2}{\left\|\left(\Lambda_z|\partial^n\Phi,\ell\rangle\right)_{\ell'}\right\|^2}
    \label{eq:gen_ovl}
\end{equation}
for $\Phi=\phi_{1,1}$ and $t_{2,2}$ (Fig.~\ref{fig:gen}a,b). We find that most numbers are close to unity as predicted by conformality, with a maximal deviation of $7\%$. We also compute the matrix element $\langle\partial^{n'}\Phi,\ell'|\Lambda^z|\partial^n\Phi,\ell\rangle$ and compare it with the value in CFT (Fig.~\ref{fig:gen}c,d). The agreement is good up to the second-order descendants. Higher descendants are subjected to larger deviations because the states receive corrections due to the irrelevant perturbations and may be mixed with states in the same symmetry sectors with similar energy. 

\begin{figure}[t]
    \centering
    \includegraphics[width=\figurewidth]{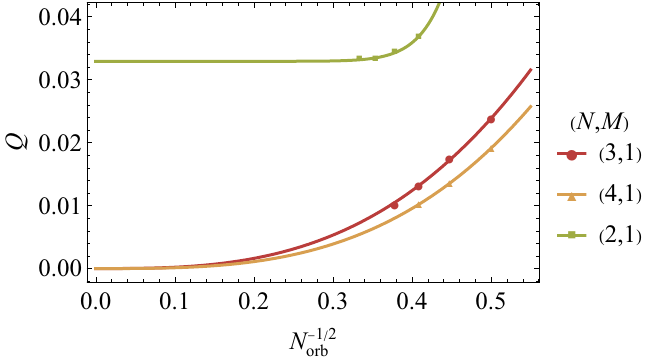}
    \caption{The minimal cost function $Q_{\min}$ as a function of system size $\Norb$ in $(N,M)=(3,1),(4,1),(2,1)$ models. The solid lines, fitted with the ansatz $Q(\Norb)=Q_0+Q_1\Norb^{-\alpha}(Q_0>0)$, are just for an eye guidance.}
    \label{fig:cost}
\end{figure}

To confirm whether or not this point describes a genuine CFT, we study how the quality of conformal symmetry scales with system size. In the $(N,M)=(3,1)$ and $(4,1)$ models, the minimal cost function $Q_{\min}$ decreases rapidly as the system size increases (Fig.~\ref{fig:cost}) and shows a trend of scaling to 0 in the thermodynamic limit, indicating perfect conformal symmetry. In contrast, $Q_{\min}$ has little size dependence in the $(N,M)=(2,1)$ model. This may suggest that the nature of the conformal region of $(3,1)$ and $(4,1)$ is qualitatively different from $(2,1)$: the $(3,1)$ and $(4,1)$ models are likely to describe genuine CFTs, while the $(2,1)$ model that describes the $\SO(5)$ DQCP is likely to be pseudocritical, although due to the size limit of our calculation, it is difficult to rule out the possibilities that $(3,1)$ and $(4,1)$ models could be pseudocritical or the $(2,1)$ model could be geniunely critical at the current stage.

We add that for the case of $N=2M$, the model has an extra particle-hole symmetry in the UV, and it becomes the parity symmetry of the CFT in the IR~\cite{Zhou2024SO5}, so one would naturally expect the parity symmetry to be broken in the CFT if $N\neq 2M$. Nevertheless, to conclude if this is true, one needs to examine the existence of parity-odd structures in the CFT correlators of spinning operators, \textit{e.g.}, three-point correlators of conserved current $\langle J^\mu J^\nu J^\rho\rangle$. We leave this for future exploration. 

\begin{table}[t]
    \centering
    \caption{The scaling dimension and quantum numbers for the low lying primary operators obtained from state-operator correspondence at different $(N,M)$. For the case of $M=N/2$, the theory has an extra parity symmetry $\mathcal{P}$. The reported numbers are raw data at the optimal point and the maximal size. Since they may receive correction from irrelevant perturbations, the value in the CFT may differ from the raw data. The scaling dimensions are calibrated to minimise the cost function, which is different from our previous paper~\cite{Zhou2024SO5}. We believe the $(2,1)$ theory is not truely conformal. The full spectrum can be found in Appendix~\ref{app:full_spec}.}
    \label{tbl:spec}
    \begin{tabular}{ccc|cc|ccc}
        \hline\hline
        op & spin & rep & \multicolumn{4}{c}{$\Delta$} &$\mathcal{P}$\\
        & & & $(3,1)$ & $(4,1)$ & $(2,1)$ & $(4,2)$& \\
        \hline 
        $\phi_{(1,1)}$ & $0$ & $\ydiagram{1,1}$               & $0.770$ & $0.795$ & $0.702$ & $0.782$ & $-$ \\
        $t_{(2,2)}$ & $0$ & $\ydiagram{2,2}$                  & $1.818$ & $1.798$ & $1.787$ & $1.767$ & $+$ \\ 
        $J^\mu_{(2)}$ & $1$ & $\ydiagram{2}$                  & $2.005$ & $2.008$ & $2.029$ & $2.002$ & $+$ \\
        $J^{+,\mu}_{(2,1,1)}$ & $1$ & $\ydiagram{2,1,1}$      & $2.521$ & $2.586$ & /       & $2.554$ & $+$ \\
        $J^{-,\mu}_{(2,1,1)}$ & $1$ & $\ydiagram{2,1,1}$      &    /    &    /    &    /    & $2.563$ & $-$ \\
        $S$ & $0$ & $\bullet$                                 & $2.979$ & $2.987$ & $2.797$ & $3.028$ & $+$ \\
        $T^{\mu\nu}$ & $2$ & $\bullet$                        & $2.998$ & $3.000$ & $2.956$ & $2.992$ & $+$ \\
        $m_{(3,3)}$ & $0$ & $\ydiagram{3,3}$                  & $3.079$ & $2.959$ & $3.170$ & $2.916$ & $+$ \\
        $J'^\mu_{(3,1)}$ & $1$ & $\ydiagram{3,1}$             & $3.269$ & $3.204$ & $3.345$ & $3.165$ & $-$ \\
        $t_{(1,1,1,1)}$ & $0$ & $\ydiagram{1,1,1,1}$          &    /    & $3.408$ &    /    & $1.172$ & $+$ \\
        $t_{(2,2,1,1)}$ & $0$ & $\ydiagram{2,2,1,1}$          &    /    & $4.144$ &    /    & $2.160$ & $-$ \\
        $S'$ & $0$ & $\bullet$                               & $4.584$ & $4.428$ & $4.370$ & $4.378$ & $-$ \\
        \hline\hline
    \end{tabular}
\end{table}

We then take a closer look at the operator spectra in the $\Sp(N)$ CFTs. The low-lying primary operators are reported in Table~\ref{tbl:spec}. The reported numbers are raw data at the optimal point and the maximal size. Since they may receive correction from irrelevant perturbations, the value in the CFT may differ from our raw data. The parameter dependence of these scaling dimensions and the full spectrum can be found in Appendices~\ref{app:dim} and \ref{app:full_spec}. In the following, we explain some notable primaries.
\begin{enumerate}
    \item $\phi_{(1,1)}$ is a traceless antisymmetric rank-2 $\Sp(N)$ tensor $\ydiagram{1,1}$. In the proximate ordered phase with symmetry-breaking into target space $\Sp(N)/(\Sp(N-M)\times\Sp(M))$, this operator serves as an order parameter. 
    \item $t_{(2,2)}$ is a rank-4 tensor. This operator can break the global symmetry to $\SU(N)\times\rU(1)$. 
    \item The global symmetry current $J_{(2)}^\mu$ is a spin-$1$ symmetric rank-2 tensor $\ydiagram{2}$ with exactly $\Delta=2$.
    \item $S$ is the lowest symmetry singlet. We find it slightly relevant in both $(N,M)=(3,1)$ and $(4,1)$ theories, which may suggest that this CFT describes a phase transition, and slightly irrelevant in the $(4,2)$ theory. We also need to note that the scaling dimension is subject to correction from irrelevant operators, given $\Delta_S$ is very close to $3$, its relevance remains to be settled. In the half-filled models, it is parity-even.
    \item $J^{\pm,\mu}_{(2,1,1)}$ are spin-$1$ rank-4 tensors $\ydiagram{2,1,1}$. This operator is new in $(N,M)=(3,1)$ and $(4,1)$ theories and does not have correspondence in the $(N,M)=(2,1)$ theory. For the $(N,M)=(4,2)$ theory, two operators of such kind with close scaling dimensions exist, respectively parity-odd and parity-even.
    \item The stress tensor $T^{\mu\nu}$ is a spin-$2$ singlet with exactly $\Delta=3$.
    \item $S'$ is the second lowest symmetry singlet. In the $N=2M$ theories, it is parity-odd and corresponds to mass $\bar{\psi}\psi$ in the QCD description.
    \item $t_{(1,1,1,1)}$ and $t_{(2,2,1,1)}$ are higher-rank tensors that exist only at $N=4$. Their scaling dimensions are high in $(4,1)$ theory but fall to very low in $(4,2)$ theory.
\end{enumerate}
We also note the following points:
\begin{itemize}
    \item The global symmetry of the CFT is actually $\Sp(N)/\mathbb{Z}_2$, as only operators in the real but not pseudoreal representations appear in the operator spectrum.
    \item The scaling dimensions for the $(N,M)=(2,1)$ theory are different from our former work~\cite{Zhou2024SO5}, it is either due to the dimension drift from the pseudocriticality, or we have found a distinct fixed point in the phase diagram. It would be interesting to study how this result fits into the widely studied $\SO(5)$ DQCP~\cite{Sandvik2007DQCP,Melko2008Scaling,Lou2009Antiferromagnetic,Chen2013Deconfined,Sumiran2013Neel,Sreejith2014Monopoles,Nahum2015DQCP,Shao2016Quantum,Takahashi2024SO5Multicriticality}.
\end{itemize}  

\textit{Candidate theories.} ---
Having presented the numerical observations of the new series of CFTs with $\Sp(N)$ symmetry, we now discuss the possible candidates for the CFTs that could theoretically exist in the phase diagram of our model. One way to answer this question is to consider the possible UV completion of the possible symmetry-broken phases. The simplest symmetry-broken phase would be filling the first $2M$ flavours out of the $2N$ flavours. It has a residual symmetry of $\Sp(M)\times\Sp(N-M)$ describing the rotation within the filled flavours and the empty flavours. Hence, it is described by the NL$\sigma$M living on a Grassmannian
\begin{equation*}
    \mathcal{G}=\frac{\Sp(N)}{\Sp(M)\times\Sp(N-M)}.
\end{equation*}
Furthermore, there is a Wess-Zumino-Witten (WZW) term with level $k=1$, as a natural generalisation of the scenario of $(N,M)=(2,1)$~\cite{Zhou2024SO5}. This symmetry-broken phase can be considered as a multi-flavour generalisation of quantum Hall ferromagnet, and the existence of WZW term directly follows a well-established result of quantum Hall ferromagnet, namely its Skyrmion carries original fermion charge~\cite{Sondhi1993Skyrmion,Moon1995Vortices,Lee2015WZW}. 

\begin{figure}[b]
    \centering
    \includegraphics[width=\figurewidth]{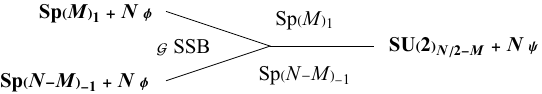}
    \caption{A phase diagram of the three candidate critical Chern-Simons-matter theories and the gapped and symmetry-broken phases in their vicinity. }
    \label{fig:pt_illus}
\end{figure}

As discussed in Ref.~\cite{Komargodski2018QCD}, one can write down several UV Lagrangians that share the same global symmetry and anomaly as this WZW-NL$\sigma$M, hence serve as the natural candidates for the $\Sp(N)$ CFTs we have discovered,
\begin{enumerate}
    \item $\Sp(M)_1+N\phi$: $N$ flavours of critical scalar fields (in the fundamental representation) coupled to an $\Sp(M)_1$ Chern-Simons field.
    \item $\Sp(N-M)_{-1}+N\phi$: $N$ flavours of critical scalar fields (in the fundamental representation) coupled to an $\Sp(N-M)_{-1}$ Chern-Simons field.
    \item $\SU(2)_{N/2-M}+N\psi$: $N$ flavours of critical fermions (in the fundamental representation) coupled to an $\Sp(1)_{N/2-M}\cong\SU(2)_{N/2-M}$ Chern-Simons field.
\end{enumerate}
For a more detailed discussion, see Appendix~\ref{app:csm}.

Semi-classically, all three theories describe phase transitions of topological orders, where the $\Sp(N)$ singlet mass of bosons or fermions is the tuning operator for the transition. Specifically, $\Sp(M)_1+N\phi$ describes the transition between the $\Sp(M)_1$ Chern-Simons theory and the symmetry breaking phase; $\Sp(N-M)_{-1}+N\phi$ describes the transition between the $\Sp(N-M)_{-1}$ Chern-Simons theory and the symmetry breaking phase; $\SU(2)_{N/2-M}+N\psi$ describes the transition between the $\Sp(M)_1$ and $\Sp(N-M)_{-1}$ Chern-Simons theories~(Fig.~\ref{fig:pt_illus}). For the case of $N=2M$, $\SU(2)_{N/2-M}+N\psi$ describes a stable critical phase rather than a phase transition because it has an extra parity symmetry that forbids the fermion mass term. We also note that $\Sp(1)_1 \cong \SU(2)_1$ is the familiar $\nu=1/2$ bosonic Laughlin state (semion topological order), while $\Sp(M)_1$ for $M>1$ corresponds to non-Abelian topological orders. In particular, $\Sp(2)_1$ and $\Sp(3)_1$ contain Ising anyon and Fibonacci anyon, respectively.  

Interestingly, our numerical data do not align well with the semi-classical expectation of any of the three candidate theories. In particular, the lowest singlet in Table~\ref{tbl:spec} is very close to $3$, which is in strong contrast to the semi-classical expectation that all three theories will have a relevant singlet (\textit{i.e.}, mass term) with scaling dimension $\Delta\approx 2$, serving as a tuning operator for the phase transition. Indeed, because of the almost marginal lowest singlet, it is difficult to determine whether our discovered CFTs describe a phase transition or a stable conformal phase. Some finite-size scaling results away from the conformal fixed point is given in Appendix~\ref{app:pt}. Clearer evidence requires a more careful analysis by conformal perturbation theory~\cite{Lao2023Perturbation}. Moreover, all three candidate theories are proximate to topological orders, but we have not found any clear signature of topological order in our phase diagram at accessible system sizes, including chiral edge modes and topological degeneracy on the torus. However, we remark that these tensions do not necessarily mean that our discovered CFTs are not one of the three Chern-Simons-matter theories, since the semi-classical expectations may not be valid at small $N$ and $M$ we are studying here.

\emph{Discussion.} --- We have presented clear evidence for the emergence of a series of CFTs in the $\Sp(N)$ Grassmannian WZW-NL$\sigma$M on the fuzzy sphere. The phase diagram of our $\Sp(N)$ model can contain a wealth of CFT candidates, including three different critical non-Abelian Chern-Simons-matter theories. It will be interesting to determine if our discovered CFTs are one of the three Chern-Simons-matter theories, and to explore a larger phase diagram to find all the three Chern-Simons-matter theories in the model. It is also intriguing to explore the topological orders in the model, particularly since most of the candidate topological orders will be non-Abelian. In addition to the bulk properties, it is also appealing to study the conformal defects and boundaries of these CFTs and explore the consequences of the anomalies. At last, extending our study to other NL$\sigma$M~\cite{Komargodski2018QCD,Read1989SpinPeierlsB,Xu2013NLSM,Zou2021Stiefel,Bi2016Stable} may lead to many fruitful results with important applications to quantum criticality and critical spin liquid.

\emph{Acknowledgement.} --- We would like to thank Chong Wang, Max Metlitski, Weicheng Ye, Ryan Lanzetta, and Davide Gaiotto for illuminating discussions. Z.Z. acknowledges support from the Natural Sciences and Engineering Research Council of Canada (NSERC) through Discovery Grants. Research at Perimeter Institute is supported in part by the Government of Canada through the Department of Innovation, Science and Industry Canada and by the Province of Ontario through the Ministry of Colleges and Universities.

\onecolumngrid
\renewcommand{\figurewidth}{0.5\linewidth}
\setlength{\tabcolsep}{10pt}
\appendix

\section{Model construction}
\label{app:model}

In this section, we explain how to express the real-space Hamiltonian Eq.~\eqref{eq:hmt} into an orbital-space Hamiltonian in terms of the LLL creation/annihilation operators $c^{(\dagger)}_{ma}$.

By substituting $\psi_a(\mathbf{r})=\sum_{m=-s}^sY_{sm}^{(s)}c_{ma}$, we obtain
\begin{multline}
    H=\sum_{m_1m_2m_3m_4}U_{m_1m_2m_3m_4}\delta_{m_1+m_2,m_3+m_4}c^\dagger_{m_1a}c^\dagger_{m_2b}c_{m_3b}c_{m_4a}\\-\frac{1}{2}\sum_{m_1m_2m_3m_4}V_{m_1m_2m_3m_4}\delta_{m_1+m_2,m_3+m_4}\Omega_{aa'}\Omega_{bb'}c^\dagger_{m_1a}c^\dagger_{m_2a'}c_{m_3b'}c_{m_4b}
\end{multline}
where
\begin{equation}
    U_{m_1m_2m_3m_4}=\int\rd^2\br_1\rd^2\br_2U(\br_{12})\bY_{sm_1}^{(s)}(\br_1)\bY_{sm_2}^{(s)}(\br_2)Y_{sm_3}^{(s)}(\br_2)Y_{sm_4}^{(s)}(\br_1).
\end{equation}
By parametrising the the potential function $U(\br_{12})$ by pseudopotentials $U_l(l=0,\dots,2s)$,
\begin{equation}
    U_{m_1m_2m_3m_4}=\sum_lU_l(4s-2l+1)\begin{pmatrix}
        s&s&2s-l\\m_1&m_2&-m_1-m_2
    \end{pmatrix}\begin{pmatrix}
        s&s&2s-l\\m_4&m_3&-m_4-m_3
    \end{pmatrix},
\end{equation}
and $V_{m_1m_2m_3m_4}$ can be expressed by a similar expression. 

The pseudopotentials can be thought of in the following way: To find out all the four-fermion interactions allowed by the rotation symmetry $\SO(3)$ and global symmetry $\Sp(N)$, we classify all the fermion bilinears $c_{m_1a}c_{m_2b}$ into irreducible representations (irrep) of $\SO(3)\times\Sp(N)$. For each irrep, by contracting the bilinear with its Hermitian conjugate, we obtain an allowed four-fermion interaction term. 

Each fermion carries $\SO(3)$ spin-$s$ and $\Sp(N)$ fundamental. For the rotation symmetry $\SO(3)$, the bilinear can carry spin-$(2s-l)(l=0,\dots,2s)$ represetation; for even $l$, the orbital indices are symmetrised; for odd $l$, the orbital indices are antisymmetrised. For the global symmetry $\Sp(N)$, the bilinear can carry singlet $\mathrm{\bullet}$, traceless antisymmetric rank-2 tensor $\ydiagram{1,1}$ and symmetric rank-2 tensor $\ydiagram{2}$ representation; for $\bullet$ and $\ydiagram{1,1}$, the flavour indices are antisymmetrised; for $\ydiagram{2}$, the flavour indices are symmetrised. As the two fermions altogether should be antisymmetrised, the allowed combinations are 
\begin{enumerate}
    \item $\Sp(N)$ singlet and $\SO(3)$ spin-$(2s-l)$ with even $l$, the bilinears are 
    \begin{equation}
        \Delta_{lm}=\sqrt{4s-2l+1}\begin{pmatrix}
            s&s&2s-l\\m_1&m_2&-m_1-m_2
        \end{pmatrix}\Omega_{aa'}c_{m_1a}c_{m_2a'}\delta_{m,m_1+m_2}
    \end{equation}
    where the prefactor $\sqrt{4s-2l+1}$ comes from the conversion between Wigner-$3j$ symbol and Clebsch-Gordan coefficients. The corresponding interaction term $H_{\mathrm{S},l}=\sum_m\Delta_{lm}^\dagger\Delta_{lm}$ is the even-$l$ pseudopotential for the $V$-term.
    \item $\Sp(N)$ antisymmetric and $\SO(3)$ spin-$(2s-l)$ with even $l$, the bilinears are 
    \begin{equation}
        \Delta_{lm,[ab]}=\sqrt{4s-2l+1}\begin{pmatrix}
            s&s&2s-l\\m_1&m_2&-m_1-m_2
        \end{pmatrix}\left(c_{m_1a}c_{m_2b}-c_{m_1b}c_{m_2a}-\tfrac{1}{N}\Omega_{ab}\Omega_{cc'}c_{m_1c'}c_{m_2c}\right)\delta_{m,m_1+m_2}.
    \end{equation}
    The corresponding interaction term $H_{\mathrm{A},l}=\sum_m\Delta_{lm,[ab]}^\dagger\Delta_{lm,[ab]}$ is the even-$l$ pseudopotential for the $U$-term.
    \item $\Sp(N)$ symmetric and $\SO(3)$ spin-$(2s-l)$ with odd $l$, the bilinears are 
    \begin{equation}
        \Delta_{lm,(ab)}=\sqrt{4s-2l+1}\begin{pmatrix}
            s&s&2s-l\\m_1&m_2&-m_1-m_2
        \end{pmatrix}\left(c_{m_1a}c_{m_2b}+c_{m_1b}c_{m_2a}\right)\delta_{m,m_1+m_2}.
    \end{equation}
    The corresponding interaction term $H_{\rT,l}=\sum_m\Delta_{lm,(ab)}^\dagger\Delta_{lm,(ab)}$ is the odd-$l$ pseudopotential for the $U$-term.
\end{enumerate}

In summary, all allowed interactions are the $U_l$ terms with both even and odd $l$, and the $V_l$ terms with only even $l$.

\FloatBarrier

\section{Technical details of the ED calculation}
\label{app:l2c2} 

In this section, we explain some technical details of the ED calculation. More details can be found in the documentation of the FuzzifiED package~\cite{FuzzifiED}. A sample code can be found at \href{https://github.com/mankai-chow/FuzzifiED.jl/blob/main/examples/sp3_spectrum.jl}{examples/sp3\_spectrum.jl} in the GitHub repository.

The $\rU(1)$ and $\mathbb{Z}_2$ subgroups of the rotation and global symmetry are implemented to divide the Hilbert space into sectors. We implement the following three kinds of conserved quantities
\begin{align}
    N_e&=\sum_{ma}c^\dagger_{ma}c_{ma}\nonumber\\
    L_z&=\sum_{ma}mc^\dagger_{ma}c_{ma}\nonumber\\
    S_{z,a}&=\sum_{m}(c^\dagger_{ma}c_{ma}-c^\dagger_{m,a+N/2}c_{m,a+N/2})&a&=1,\dots,N.
\end{align}
To go through all the symmetry representations, it suffices to consider only the $L_z=0,S_{z,a}=0$ sector. We implement the following three kinds of $\mathbb{Z}_2$ symmetries, 
\begin{align}
    \mathcal{R}_y:c_{ma}&\mapsto(-1)^{s-m}c_{-m,a}\nonumber\\
    \mathcal{Z}_a:c_{ma}&\mapsto c_{m,a+N/2}&c_{m,a+N/2}&\mapsto-c_{ma}&a=1,\dots,N\nonumber\\
    \mathcal{X}_1:c_{m1}&\leftrightarrow c_{m2}&c_{m,1+N/2}&\leftrightarrow c_{m,2+N/2}.
\end{align}
To go through all the symmetry representations, it suffices to consider only $(\mathcal{Z}_1,\mathcal{Z}_2,\mathcal{Z}_3)=(+,+,+),\allowbreak(+,+,-),\allowbreak(-,-,+),\allowbreak(-,-,-)$ sectors for the $(3,1)$ model and $(\mathcal{Z}_1,\mathcal{Z}_2,\mathcal{Z}_3,\mathcal{Z}_4)=(+,+,+,+),\allowbreak(+,+,+,-),\allowbreak(+,+,-,-),\allowbreak(-,-,-,+),\allowbreak(-,-,-,-)$ sectors for the $(4,1)$ model and $\mathcal{R}_y=\pm,\mathcal{X}_1=\pm$.

To determine the representations of each state under $\SO(3)$ rotation and $\Sp(N)$ symmetry, we measure the total angular momentum $L^2$ and quadratic Casimir $C_2$. The $L^2$ operator can be constructed using $L_{z,\pm}$ in the following way
\begin{align}
    L_z&=\sum_{ma}mc^\dagger_{ma}c_{ma}\nonumber\\
    L_\pm&=\sum_{ma}\sqrt{(s\mp m)(s\pm m+1)}c^\dagger_{m\pm 1,a}c_{ma}\nonumber\\
    L^2&=L_+L_-+L_z^2-L_z.
\end{align}
Its expectation value is related to the spin of the state by $\langle\Phi|L^2|\Phi\rangle=\ell_\Phi(\ell_\Phi+1)$
The Casimir $C_2$ can be constructed using the generators $t_{(ab)}$ of the $\Sp(N)$ group
\begin{align}
    t_{(ab)}&=\frac{i}{2}\sum_{m}\left(\Omega_{ac}c^\dagger_{mc}c_{mb}+\Omega_{bc}c^\dagger_{mc}c_{ma}\right)\nonumber\\
    C_2&=\Omega_{aa'}\Omega_{bb'}t_{(a'b')}t_{(ab)}.  
\end{align}
The correspondence between its expectation value and the $\Sp(N)$ representation is listed in Table~\ref{tbl:rep}.

\begin{table}[htbp]
    \centering
    \caption{The dimension, quadratic Casimir $C_2$ and Young diagram of the representations of $\Sp(N)$ group for $N=2,3,4$.}
    \label{tbl:rep}
    \begin{tabular}{c|cc|cc|cc}
        \hline\hline
        &\multicolumn{2}{c|}{$\Sp(2)$}&\multicolumn{2}{c|}{$\Sp(3)$}&\multicolumn{2}{c}{$\Sp(4)$}\\
        rep.&dim.&$C_2$&dim.&$C_2$&dim.&$C_2$\\
        \hline 
        $\bullet$         &  $1$ & $0$ &   $1$ &  $0$ &   $1$ &  $0$ \\
        $\ydiagram{1,1}$  &  $5$ & $2$ &  $14$ &  $3$ &  $27$ &  $4$ \\
        $\ydiagram{2}$    & $10$ & $3$ &  $21$ &  $4$ &  $36$ &  $5$ \\
        $\ydiagram{2,2}$  & $14$ & $5$ &  $90$ &  $7$ & $308$ &  $9$ \\
        $\ydiagram{3,3}$  & $30$ & $9$ & $385$ & $12$ &$2184$ & $15$ \\
        $\ydiagram{3,1}$  & $35$ & $6$ & $189$ &  $8$ & $594$ & $10$ \\
        $\ydiagram{2,1,1}$&    / &   / &  $70$ &  $6$ & $315$ &  $8$ \\
        $\ydiagram{1,1,1,1}$&  / &   / &     / &    / &  $42$ &  $6$ \\
        $\ydiagram{2,2,1,1}$&  / &   / &     / &    / & $792$ & $11$ \\
        \hline\hline
    \end{tabular}
\end{table}

\FloatBarrier

\section{Cost function for the conformal symmetry}
\label{app:cost}

In this section, we explain the definition of the cost function. We define the cost function to capture the deviations of the scaling dimensions of the operators  $\partial^\mu\phi_{(1,1)},\allowbreak\partial^\mu\partial^\nu\phi_{(1,1)},\allowbreak\Box\phi_{(1,1)},\allowbreak\partial^\mu t_{(2,2)},\allowbreak J_{(2)}^\mu,\allowbreak\partial^\nu J_{(2)}^\mu,\allowbreak\epsilon_{\mu\nu\rho}\partial^\nu J_{(2)}^\nu$ and $T^{\mu\nu}$ from the predictions of conformal symmetry. We pick out those states according to the following criteria
\begin{itemize}
    \item $|\bbI\rangle,|\phi\rangle,|\partial\phi\rangle,|\partial\partial\phi\rangle,|t\rangle,|\partial t\rangle,|J\rangle,|\partial J\rangle$ and $|T\rangle$ as the lowest state with corresponding $L^2$ and $C_2$, and 
    \item $|\Box\phi\rangle$ and $|\epsilon\partial J\rangle$ as the second lowest state with corresponding $L^2$ and $C_2$.
\end{itemize}
We define the vectors of energy and scaling dimension differences 
\begin{align}
    \mathbf{E}&=(E_{\partial\phi}-E_\phi,E_{\partial\partial\phi}-E_\phi,E_{\Box\phi}-E_\phi,E_{\partial t}-E_t,E_J-E_{\bbI},E_{\partial J}-E_{\bbI},E_{\epsilon\partial J}-E_{\bbI},E_T-E_{\bbI})\nonumber\\
    \bDelta&=(1,2,2,1,2,3,3,3).
\end{align}
At perfect conformal symmetry, these two vectors should be proportional, 
\begin{equation*}
    \bDelta=\alpha\mathbf{E}.
\end{equation*}
The cost function is defined to be proportional to the modulus of their difference
\begin{equation*}
    Q(U_l,V_l,\alpha)=\frac{1}{\sqrt 8}\|\bDelta-\alpha\mathbf{E}\|.
\end{equation*}
Here, the cost function is a function of both the parameters $U_l,V_l$ in the Hamiltonian but also the coefficient $\alpha$. The minimisation with respect to $\alpha$ can be conducted analytically 
\begin{align}
    \alpha&=(\mathbf{E}\cdot\bDelta)/(\mathbf{E}\cdot\mathbf{E})\nonumber\\
    Q(U_l,V_l)&=\frac{1}{\sqrt 8}\left\|\bDelta-\frac{\mathbf{E}\cdot\bDelta}{\mathbf{E}\cdot\mathbf{E}}\mathbf{E}\right\|.
\end{align}
We then perform a minimisation with respect to the parameters $U_l$ and $V_l$ in the Hamiltonian to find the parameters with the best conformality using the Nelder-Mead algorithm. The parameters and minimal cost functions are listed in Table~\ref{tbl:cost}.

\begin{table}[htbp]
    \centering
    \caption{The parameters that minimise the cost function, and the minimum value of the cost function for different $(N,M)$ and system size $\Norb$.}
    \label{tbl:cost}
    \begin{tabular}{cc|cccc|c}
        \hline\hline
        $(N,M)$ & $\Norb$ & $U_1$ & $U_2$ & $V_0$ & $V_2$ & $Q_{\min}$ \\
        \hline 
        $(3,1)$ & $7$ & $\sm0.2643$ & $\sm0.0652$ & $\sm0.3798$ & $  -0.0219$ & $0.01006$ \\
                & $6$ & $\sm0.2295$ & $\sm0.0471$ & $\sm0.4003$ & $  -0.0205$ & $0.01303$ \\
                & $5$ & $\sm0.1750$ & $\sm0.0155$ & $\sm0.4328$ & $  -0.0195$ & $0.01730$ \\
                & $4$ & $\sm0.0551$ & $  -0.0718$ & $\sm0.5022$ & $  -0.0191$ & $0.02378$ \\
        \hline 
        $(4,1)$ & $6$ & $\sm0.0404$ & $  -0.0052$ & $\sm0.3934$ & $  -0.0236$ & $0.01015$ \\
                & $5$ & $  -0.0122$ & $  -0.0485$ & $\sm0.4143$ & $  -0.0258$ & $0.01350$ \\
                & $4$ & $  -0.1353$ & $  -0.1576$ & $\sm0.4649$ & $  -0.0318$ & $0.01911$ \\
        \hline 
        $(2,1)$ & $9$ & $\sm0.4561$ & $\sm0.1041$ & $\sm0.3036$ & $\sm0.0380$ & $0.03322$ \\
                & $8$ & $\sm0.4479$ & $\sm0.1047$ & $\sm0.3285$ & $\sm0.0311$ & $0.03331$ \\
                & $7$ & $\sm0.4326$ & $\sm0.1022$ & $\sm0.3601$ & $\sm0.0247$ & $0.03431$ \\
                & $6$ & $\sm0.4066$ & $\sm0.0943$ & $\sm0.4003$ & $\sm0.0193$ & $0.03680$ \\
                & $5$ & $\sm0.3786$ & $\sm0.0855$ & $\sm0.4240$ & $\sm0.0379$ & $0.04322$ \\
        \hline 
        $(4,2)$ & $5$ & $\sm0.0169$ & $  -0.0503$ & $\sm0.3718$ & $\sm0.0019$ & $0.01453$\\
                & $4$ & $\sm0.0471$ & $  -0.0502$ & $\sm0.3688$ & $\sm0.0258$ & $0.02212$ \\
        \hline\hline
    \end{tabular}
\end{table}

\FloatBarrier

\section{Conformal generators}
\label{app:gen}

The conformal generator $\Lambda^\mu=P^\mu+K^\mu$ on the states is the $l=1$ component of the Hamiltonian density, which is the local density-density interactions with some full derivatives
\begin{multline}
    \mathcal{H}(\br)=n\left(g_{U,0}+g_{U,1}\nabla^2+g_{U,2}\nabla^4\right)n+\Omega_{aa'}\Omega_{bb'}n_{ab}\left(g_{V,0}+g_{V,1}\nabla^2+g_{V_2}\nabla^4\right)n_{a'b'}\\
    +g_{D,1}\nabla^2n+g_{D,2}\nabla^2n^2+g_{D,3}\nabla^2(\Omega_{aa'}\Omega_{bb'}n_{ab}n_{a'b'})+\dots
\end{multline}
where $n=\psi_a^\dagger\psi_a,n_{ab}=\psi_a^\dagger\psi_b$, $g_{U,i}, g_{V,i}$ are linear combinations of pseudopotentials $U_i,V_i$, and $g_{D,i}$ are undetermined constants that does not affect the Hamiltonian $H=\int\rd^2\br\,\mathcal{H}$. We have only listed a few examples of the allowed full derivatives. The generator is expressed as 
\begin{equation}
    \Lambda_m=P_m+K_m=\int\rd^2\br\,Y_{l=1,m}(\br)\mathcal{H}(\br).
\end{equation}
To determine those constants, we consider another strategy by combining four fermion operators into $\SO(3)$ spin-1 and $\Sp(N)$ singlet operators. Similar to what we have done in Appendix~\ref{app:model}, we combine the fermion bilinears $\Delta_{lm},\Delta_{lm,[ab]},\Delta_{lm,(ab)}$.
\begin{multline*}
    \Lambda_m=\sum_{\substack{l_1l_2\in2\mathbb{Z}\\m_1m_2}}\tilde{V}_{l_1l_2}\Delta^\dagger_{l_1m_1}\Delta_{l_2m_2}\begin{pmatrix}
        2s-l_1&2s-l_2&1\\-m_1&m_2&m
    \end{pmatrix}+\sum_{\substack{l_1l_2\in2\mathbb{Z}\\m_1m_2}}\tilde{U}_{l_1l_2}\Delta^\dagger_{l_1m_1,[ab]}\Delta_{l_2m_2,[ab]}\begin{pmatrix}
        2s-l_1&2s-l_2&1\\-m_1&m_2&m
    \end{pmatrix}\\
    +\sum_{\substack{l_1l_2\in2\mathbb{Z}+1\\m_1m_2}}\tilde{U}_{l_1l_2}\Delta^\dagger_{l_1m_1,(ab)}\Delta_{l_2m_2,(ab)}\begin{pmatrix}
        2s-l_1&2s-l_2&1\\-m_1&m_2&m
    \end{pmatrix}.
\end{multline*}
Since $l_1-l_2\in2\mathbb{Z}$ and $|l_1-l_2|\leq 1$ in all cases, we conclude $l_1=l_2$, so
\begin{multline}
    \Lambda_m=\sum_{\substack{l\in2\mathbb{Z}\\m_1m_2}}\tilde{V}_l\Delta^\dagger_{lm_1}\Delta_{lm_2}\begin{pmatrix}
        2s-l&2s-l&1\\-m_1&m_2&m
    \end{pmatrix}+\sum_{\substack{l\in2\mathbb{Z}\\m_1m_2}}\tilde{U}_l\Delta^\dagger_{lm_1,[ab]}\Delta_{lm_2,[ab]}\begin{pmatrix}
        2s-l&2s-l&1\\-m_1&m_2&m
    \end{pmatrix}\\
    +\sum_{\substack{l\in2\mathbb{Z}+1\\m_1m_2}}\tilde{U}_l\Delta^\dagger_{lm_1,(ab)}\Delta_{lm_2,(ab)}\begin{pmatrix}
        2s-l&2s-l&1\\-m_1&m_2&m
    \end{pmatrix}.
\end{multline}
Here, $\tilde{U}_l$ and $\tilde{V}_l$ are tuning parameters. Like what we have taken in the Hamiltonian, we consider only $\tilde{U}_{0,1,2}$ and $
\tilde{V}_{0,2}$. From the algebra of the conformal generators, the $\Lambda_{0,\pm 1}$ are connected by commuting with the angular momenta
\begin{equation*}
    \Lambda_{m=\pm 1}=\frac{1}{\sqrt{2}}[\Lambda_{m=0},L_\pm].
\end{equation*}
We note that this equation holds on the level of the UV microscopic model.

Let us derive the expression for $|\Lambda\Phi,l'm'\rangle=(\Lambda^\mu|\Phi_l\rangle)_{l'm'}$. It involves the combination of angular momenta $l$ and $1$ into $l'$. Without loss of generality, we only consider the $m'=0$ component
\begin{equation}
    |\Lambda\Phi,l'0\rangle=\sum_{m=0,\pm1}\frac{(-1)^{l-l'}}{\sqrt{2l'+1}}
    \begin{pmatrix}l&1&l'\\-m&m&0\end{pmatrix}\Lambda_{m}|\Phi_{lm}\rangle.
\end{equation}
We also note that the states $\Phi_{lm}$ with different $m$ are connected by the angular momenta 
\begin{equation*}
    L_\pm|\Phi_{lm}\rangle=\sqrt{(l\mp m)(l\pm m+1)}|\Phi_{l,m\pm 1}\rangle,\qquad 
\end{equation*}
specifically, 
\begin{equation*}
    |\Phi_{l,\pm1}\rangle=\frac{1}{\sqrt{l(l+1)}}|L_\pm|\Phi_{l0}\rangle.
\end{equation*}

Hence, the $l'=l\pm 1, l$ states can be generated by acting $L_\pm,\Lambda_0$ on $|\Phi_{l0}\rangle$,
\begin{align}
    |\Lambda\Phi,(l+1)0\rangle&=\left(\sqrt{\frac{(l+1)^3}{2l+1}}\Lambda_0-\sqrt{\frac{1}{4(l+1)(2l+1)}}(L_+\Lambda_0L_-+L_-\Lambda_0L_+)\right)|\Phi_{l0}\rangle\nonumber\\
    |\Lambda\Phi,(l-1)0\rangle&=\left(\sqrt{\frac{l^3}{2l+1}}\Lambda_0-\sqrt{\frac{1}{4l(2l+1)}}(L_+\Lambda_0L_-+L_-\Lambda_0L_+)\right)|\Phi_{l0}\rangle\nonumber\\
    |\Lambda\Phi,l0\rangle&=\frac{1}{\sqrt{8l(l+1)}}(L_+\Lambda_0L_--L_-\Lambda_0L_+)|\Phi_{l0}\rangle.
\end{align}

In practice, we determine the coefficients $\tilde{U}_l$ and $\tilde{V}_l$ by minimising the modulus of $\Lambda$ acting on ground state $\|\Lambda|0\rangle\|$ and determine the overall factor by calibrating $\langle\partial^\mu j^\nu|\Lambda_0|j^\mu\rangle=2$. We supplement Fig.~\ref{fig:gen} with the results of the multiplet of $j^\mu$ in Tables~\ref{tbl:gen_st_j} and \ref{tbl:gen_ovl_j}. 

\begin{table}[htbp]
    \centering
    \caption{The matrix elements $\langle\partial^{n'}\Phi, \ell'|\left(\Lambda^\mu|\partial^n\Phi,\ell\rangle\right)_{\ell'}$ as in Eq.~\eqref{eq:gen_ovl} in the multiplet of $J^\mu$. Closer total squared overlap to unity shows better conformality.}
    \label{tbl:gen_st_j}
    \begin{tabular}{cc|cc|cc|c}
        \hline\hline
        $|\partial^n\Phi,\ell\rangle$&$\ell'$&$\langle\partial^{n'}\Phi,\ell'|$&$|\textrm{overlap}|^2$&$\langle\partial^{n'}\Phi,\ell'|$&$|\textrm{overlap}|^2$&total\\
        \hline 
        $J^\mu$&$1$&$\epsilon_{\mu\nu\rho}\partial^\nu J^\rho$&$0.9656$&/&/&$0.9656$\\
        &$2$&$\partial^\nu J^\mu$&$0.9922$&/&/&$0.9922$\\
        \hline
        $\epsilon_{\mu\nu\rho}\partial^\nu J^\rho$&$1$&$ J^\mu$&$0.3287$&$\Box J^\mu$&$0.6576$&$0.9863$\\
        &$2$&$\epsilon_{\mu\rho\sigma}\partial^\nu\partial^\rho J^\sigma$&$0.9775$&/&/&$0.9775$\\
        \hline
        $\partial^\nu J^\mu$&$1$&$ J^\mu$&$0.9031$&$\Box J^\mu$&$0.0868$&$0.9900$\\
        &$2$&$\epsilon_{\mu\rho\sigma}\partial^\nu\partial^\rho J^\sigma$&$0.9172$&/&/&$0.9172$\\
        &$3$&$\partial^\mu\partial^\nu J^\rho$&$0.9275$&/&/&$0.9275$\\
        \hline\hline
    \end{tabular}
\end{table}

\begin{table}[htbp]
    \centering
    \caption{The matrix elements $\langle\partial^{n'}\Phi, \ell'|\Lambda^z|\partial^n\Phi,\ell\rangle$ as in Fig.~\ref{fig:gen}c,d for the multiplet of $J^\mu$, compared with the value from CFT.}
    \label{tbl:gen_ovl_j}
    \begin{tabular}{cc|ccc}
        \hline\hline
        $|\partial^n\Phi,\ell\rangle$&$\langle\partial^{n'}\Phi, \ell'|$&fuzzy sphere&CFT&error\\
        \hline
        $J^\mu$&$\partial^\nu J^\mu$&$2.0000$&$2.0000$&/\\
        $\partial^\nu J^\mu$&$\Box J^\mu$&$0.6202$&/&/\\
        $\partial^\nu J^\mu$&$\partial^\mu\partial^\nu J^\rho$&$2.7356$&$3.0984$&$12\%$\\
        $\epsilon_{\mu\nu\rho}\partial^\nu J^\rho$&$\epsilon_{\mu\rho\sigma}\partial^\nu\partial^\rho J^\sigma$&$2.0957$&$2.4495$&$14\%$\\
        \hline\hline
    \end{tabular}
\end{table}

\FloatBarrier

\section{Candidate theories for the WZW-NLSM}
\label{app:csm}

\newcommand{\bQ}{\mathbf{Q}}
\newcommand{\bU}{\mathbf{U}}
\newcommand{\bg}{\mathbf{g}}
\newcommand{\tbQ}{\tilde{\mathbf{Q}}}
\newcommand{\tr}{\operatorname{tr}}
In this appendix, we discuss the derivation of the non-linear sigma model (NL$\sigma$M) and the possible renormalisable theories that can live on its phase diagrams. 

We first explain why Eq.~\eqref{eq:hmt} gives an $\Sp(2N)$ Grassmannian NL$\sigma$M with a level-1 WZW term. The $N=2$ case has been discussed in Refs.~\cite{Lee2015WZW,Ippoliti2018TorusSO5}. When $V(\br)=0$, the dynamics of the system is captured by a NL$\sigma$M on the $\rU(2N)$ Grassmannian $\rU(2N)/(\rU(2M)\times \rU(2(N-M)))$,
\begin{equation} 
    S[\bQ] = \frac{1}{g} \int \rd^2 \br\,\rd\tau\tr(\partial^\mu\bQ(\br, t))^2 + S_{\textrm{WZW},1}[\bQ].
\label{eq:NLSM}
\end{equation}
Here $\bQ(\br, t)$ is a $2N\times 2N$ matrix field living on the $\rU(2N)$ Grassmannian, parameterised by
\begin{equation}
    \bQ = \bg\begin{pmatrix} \bbI_{2M} & 0 \\ 0 & -\bbI_{2(N-M)} \end{pmatrix}\bg^{-1},\quad \bg\in\rU(2N),\quad \bg^{-1}=\bg^\dagger
\end{equation}
with $\bg$ being a $\rU(2N)$ matrix. The matrix field $\bQ(\br, t)$ encodes the occupation of fermions in our original system, specifically describing which $2M$ fermions out of the total $2N$ are occupied at the space-time coordinates $(\br, t)$. The WZW term reads 
\begin{equation}
    S_{\textrm{WZW},k}[\bQ]=\int_0^1\rd u\int\rd^2\br\,\rd\tau\frac{2\pi ik}{(16\pi)^2}\epsilon^{\mu\nu\rho\sigma}\tr(\tbQ\partial_\mu\tbQ\partial_\nu\tbQ\partial_\rho\tbQ\partial_\sigma\tbQ)
\end{equation}
where $\tbQ(\br,\tau,u)$ is an extension of $\bQ(\br,\tau)$ into an auxiliary fourth dimension parametrised by $0\leq u\leq 1$ with boundary condition 
\begin{equation*}
    \tbQ(\br,\tau,u=1)=\bQ(\br,\tau),\quad \tbQ(\br,\tau,u=0)=\begin{pmatrix} \bbI_{2M} & 0 \\ 0 & -\bbI_{2(N-M)} \end{pmatrix}
\end{equation*}
It has a simple physical interpretation: the skyrmion is a fermion carrying a $\rU(1)$ electronic charge. This generalises a well-established result of the quantum Hall ferromagnet~\cite{Sondhi1993Skyrmion}, corresponding to $M=N=1/2$ in our scenario. Specifically, one can consider a special skyrmion that exhibits non-trivial patterns solely in the first two flavours of fermions, which then reduces to the old story of the quantum Hall ferromagnet. 

Introducing a finite $V(\br)$, the global $\mathrm{SU}(2N)$ symmetry is explicitly broken down to the $\Sp(N)$ symmetry. As a consequence, the matrix field $\bQ(\br, t)$ becomes energetically favourable to fluctuate on the $\Sp(N)$ Grassmannian $\mathcal{G}=\Sp(N)/(\Sp(M)\times \Sp(N-M))$, which is a submanifold of the $\rU(2N)$ Grassmannian. Additionally, the WZW term defined on the $\rU(2N)$ Grassmannian is reduced to a WZW term on the $\Sp(N)$ Grassmannian. Therefore, at finite $V(\br)$ the system can be effectively described by Eq.~\eqref{eq:NLSM}, where the matrix field $\bQ(\br, t)$ resides on the $\Sp(N)$ Grassmannian and is parameterised by
\begin{equation}
    \bQ = \bg\begin{pmatrix} \bbI_{2M} & 0 \\ 0 & -\bbI_{2(N-M)} \end{pmatrix}\bg^{-1},\quad \bg\in\Sp(N),\quad \bg^{-1}=-\bOmega_N\bg^\rT\bOmega_N
    \label{eq:grassmannianmatrix}
\end{equation}
The stiffness of the NL$\sigma$M depends on the interactions $V(\br)$ and $U(\br)$ in our model Hamiltonian. The WZW term has a similar expression
\begin{equation}
    S_{\textrm{WZW},k}[\bQ]=\int_0^1\rd u\int\rd^2\br\,\rd\tau\frac{2\pi ik}{(16\pi)^2}\epsilon^{\mu\nu\rho\sigma}\tr(\tbQ\partial_\mu\tbQ\partial_\nu\tbQ\partial_\rho\tbQ\partial_\sigma\tbQ)
\end{equation}
where $\tbQ(\br,\tau,u)$ is an extension of $\bQ(\br,\tau)$ into an auxiliary fourth dimension parametrised by $0\leq u\leq 1$ with boundary condition 
\begin{equation*}
    \tbQ(\br,\tau,u=1)=\bQ(\br,\tau),\quad \tbQ(\br,\tau,u=0)=\begin{pmatrix} \bbI_{2M} & 0 \\ 0 & -\bbI_{2(N-M)} \end{pmatrix}
\end{equation*}

We then discuss how the above WZW-NL$\sigma$M is related to the bosonic Chern-Simons-matter theory of $N$ flavours of critical scalar fields coupled to an $\Sp(M)_1$ or $\Sp(N-M)_{-1}$ Chern-Simons field. For the case of $\Sp(M)_1$, one can start with the Stiefel manifold $\Sp(N)/\Sp(N-M)$. It can be parametrised by a $2N\times 2M$ matrix $\mathbf{n}$ such that $\mathbf{n}^\rT\bOmega_N\mathbf{n}=\bOmega_{M}$. We can interpret $\mathbf{n}$ as a matrix field $\phi_{ia}$ that is bi-fundamental of $\Sp(N)$ and $\Sp(M)$, $i=1,\cdots, 2N$ and $a=1,\cdots,2M$. To recover the original Grassmannian $\mathcal{G}$, we now gauge the $\Sp(M)$ symmetry of the Stiefel manifold. In the language of matrix field $\phi_{ia}$, this gauging process corresponds to couple it an $\Sp(M)$ gauge field. So we end up with a theory with $N$ flavours of bosonic field coupled to an $\Sp(M)$ gauge field, and the bosonic field is in the fundamental representation of the $\Sp(M)$ gauge group. The WZW term of the original NL$\sigma$M corresponds to the Chern-Simons term of the gauge field $\Sp(M)$~\cite{Komargodski2018QCD}, and the WZW level is identical to the Chern-Simons level. We therefore arrive at $N$ flavours of critical scalar fields coupled to an $\Sp(M)_1$ Chern-Simons field
\begin{equation}
    \mathscr{L}[\phi^{ia},\sigma,A_\mu]=(D_{A,\mu}\phi)^{ia}\Omega_{ij}\Omega_{ab}(D_A^\mu\phi)^{jb}+\sigma\phi^{ia}\Omega_{ij}\Omega_{ab}\phi^{jb}-\frac{N}{2\lambda}\sigma^2+\mathcal{L}_{\textrm{CS},1}[A_\mu]
\end{equation}
where $i,j=1,\dots,2N$ are the $\Sp(N)$ flavour indices, $a,b=1,\dots,2M$ are the $\Sp(M)$ gauge indices, $A_\mu$ is the $\Sp(M)$ gauge connection. The covariant derivative $D_{A,\mu}=\partial_\mu-iA_\mu$. The level-$k$ Chern-Simons term is 
\begin{equation}
    \mathcal{L}_{\textrm{CS},k}[A_\mu]=\frac{k}{4\pi}\epsilon^{\mu\nu\rho}\tr(A_\mu\partial_\nu A_\rho+\tfrac{2}{3}A_\mu A_\nu A_\rho),
\end{equation}
where the trace is taken with respect to the $\Sp(M)$ gauge index.

Similarly, if we start from the Stiefel manifold $\Sp(N)/\Sp(M)$ and couple it to an $\Sp(N-M)_{-1}$ gauge field, we will get $N$ flavours of critical scalar fields coupled to the $\Sp(N-M)_{-1}$ Chern-Simons field. 
\begin{equation}
    \mathscr{L}[\phi^{ia},\sigma,A_\mu]=(D_{A,\mu}\phi)^{ia}\Omega_{ij}\Omega_{ab}(D_A^\mu\phi)^{jb}+\sigma\phi^{ia}\Omega_{ij}\Omega_{ab}\phi^{jb}-\frac{N}{2\lambda}\sigma^2+\mathcal{L}_{\textrm{CS},-1}[A_\mu]
\end{equation}
where the flavour index $i,j=1,\dots,2N$ and the gauge index $a,b=1,\dots,2(N-M)$.

Next, we show that the WZW-NL$\sigma$M is related to a fermionic theory, namely $N$ flavours of fermions (in the fundamental representation) coupled to the $\Sp(1)_{N/2-M}\cong\SU(2)_{N/2-M}$ Chern-Simons field, whose Lagrangian reads
\begin{equation}
    \mathscr{L}[\psi^{ia},A_\mu]=\psi^{ias_1}\Omega_{ij}\Omega_{ab}\Omega_{s_1s'}i(\gamma^\mu)^{s'}{}_{s_2}(D_{A,\mu}\psi)^{jbs_2}+\mathcal{L}_{\textrm{CS},M-N/2}[A_\mu]
\end{equation}
where the $\Sp(N)$ flavour index $i,j=1,\dots,N$, the $\Sp(1)\cong \SU(2)$ gauge index $a,b=1,2$, and the $\Sp(1)\cong \SU(2)$ spinor index $s=1,2$.

We first couple the fermions to a bosonic mass field $\mathbf{Q}$ that lives in the $\Sp(N)$ Grassmannian $\mathcal{G}$. The coupling reads $\tr(\bar{\bPsi}\mathbf{Q}\bPsi)=\psi^{ias}\Omega_{ii'}\Omega_{ab}\Omega_{ss'}Q^{i'}{}_j\psi^{jbs'}$, and $\bPsi$ and $\bar{\bPsi}$ are matrix fields defined as $\Psi^{i,as}=\psi^{ias}$, $\bar{\Psi}_{bs',i'}=\psi^{ias}\Omega_{ii'}\Omega_{ab} \Omega_{ss'}$. Since $\mathbf{Q}$ lives in the $\Sp(N)$ Grassmannian $\mathcal{G}$ written as Eq.~\eqref{eq:grassmannianmatrix}, we can rewrite the coupling as 
\begin{equation*}
    \tr(\bar{\bPsi}\mathbf{Q}\bPsi)=\tr\left(\bar{\bPsi}'\begin{pmatrix}
        \bbI_{2M} & 0 \\ 0 & -\bbI_{2(N-M)}
    \end{pmatrix}{\bPsi}'\right),\quad {\bPsi}'=\bg\bPsi.  
\end{equation*}
So the mass field $\mathbf{Q}$, if condensed, will break the $\Sp(N)$ global symmetry to $\Sp(M)\times\Sp(N-M)$ global symmetries. Now we integrate out the fermions and expand the theory in terms of the mass field $\mathbf{Q}$ yielding a NL$\sigma$M on the $\Sp(N)$ Grassmannian $\mathcal{G}$, with a WZW term from integrating out fermions~\cite{Abanov2001}. Furthermore, integrating out fermions also produces a Chern-Simons term $\Sp(1)_{M-N/2}$, which exactly cancels the original Chern-Simons term in the gauge theory.
Therefore, we are left with a pure $\Sp(1)$ gauge field, which will trivially confine. We remark that in the above argument, if we are choosing the mass field $\mathbf{Q}$ to live in a different Grassmannian, we will end up with a non-trivial Chern-Simons term of the $\Sp(1)$ gauge field, which will not disappear in the IR.

At last, we discuss the phase diagram of these candidate theories. Starting from the $N$ flavours of critical scalar fields coupled to an $\Sp(M)_1$ Chern-Simons field, by gapping out the scalar fields or condensing them to Higgs the gauge field, we obtain an $\Sp(M)_1$ CS theory and a spontaneous symmetry-breaking into $\mathcal{G}$. The $\Sp(M)_1+N\phi$ theory therefore describes a phase transition between the topological ordered phase described by the $\Sp(M)_1$ CS theory and a SSB phase described by NL$\sigma$M living on $\mathcal{G}$. Similarly, the $\Sp(N-M)_{-1}+N\phi$ theory describes a phase transition between the topological orders described by the $\Sp(N-M)_{-1}$ CS theory and a SSB phase described by NL$\sigma$M living on $\mathcal{G}$. For the fermionic theory of $N$ flavours of critical fermions coupled to $\SU(2)_{N/2-M}$ Chern-Simons field, by gapping out the fermions with a mass of different signs, each fermion would add an additional level $\pm1/2$ to the $\SU(2)$ gauge field. Therefore, we arrive at $\SU(2)_{N-M}$ and $\SU(2)_{-M}$ CS theories. By level-rank duality, they are dual to the $\Sp(N-M)_{-1}$ and $\Sp(M)_1$ theories. Therefore, $\SU(2)_{N/2-M}+N\psi$ describes the phase transition between two topological orders described by $\Sp(N-M)_{-1}$ and $\Sp(M)_1$ CS theories. We therefore obtain the phase diagram illustrated in Fig.~\ref{fig:pt_illus}. We also add that these three theories are not dual to each other, as they violate the flavour bound~\cite{Hsin2016LevelRank}.

\FloatBarrier

\section{Stability of the fixed point}
\label{app:pt}

\begin{figure}[t]
    \centering
    \includegraphics[width=\figurewidth]{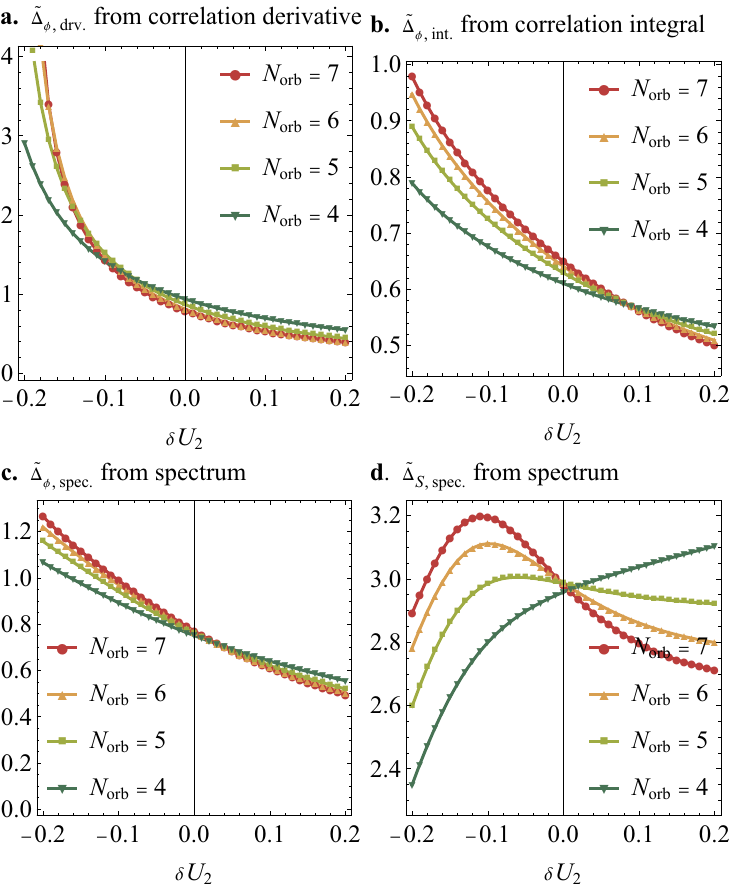}
    \caption{The dimensionless quantities (a) $\tilde{\Delta}_{\phi,\mathrm{drv.}}$ from the correlation derivative, (b) $\tilde{\Delta}_{\phi,\mathrm{int.}}$ from the correlation integral, (c) $\tilde{\Delta}_{\phi,\textrm{spec.}}$ and (d) $\tilde{S}_{\phi,\textrm{spec.}}$ defined in Eq.~\eqref{eq:dim} from the energy spectrum as a function of $\delta U_2$ on the Hamiltonian in the $(3,1)$ model at different system sizes. }
    \label{fig:fss}
\end{figure}

To investigate the stability of the CFT, \textit{i.e.}, whether it is a stable phase or a phase transition and to what phases it is proximate, we look at the behaviour of the order parameter and its two-point correlation. The order parameter is the density operator (\textit{i.e.}, fermion bilinear) in the antisymmetric rank-2 tensor representation
\begin{equation}
    n^\rA_{[ab]}(\br)=\Omega_{ac}\psi_c^\dagger(\br)\psi_b(\br)-\Omega_{bc}\psi_c^\dagger(\br)\psi_a(\br)-\tfrac{1}{N}\Omega_{ab}\psi_c^\dagger(\br)\psi_c(\br)
\end{equation}
that captures the symmetry breaking into the target space $\Sp(N)/(\Sp(N-1)\times\SU(2))$. At the conformal point, $n^\rA$ receives its lowest contribution from the operator $\phi_{(1,1)}$, and its correlation function behaves like 
\begin{equation}
    C^\rA(\br_{12})=\langle n^\rA(\br_1)n^\rA(\br_2)\rangle=\textrm{constant}\times r_{12}^{-2\Delta_\phi}
\end{equation}
where the linear and angular distance is related by $r_{12}=2\cos\theta_{12}/2$. From the correlation functions and the energy spectrum we can obtain the dimensionless quantities 
\begin{align}
    \tilde{\Delta}_{\phi,\mathrm{drv.}}&=-\left.\frac{\partial\log C^\rA(\br_{12})}{\partial r_{12}}\right|_{\theta_{12}=\pi}\nonumber\\
    \tilde{\Delta}_{\phi,\mathrm{int.}}&=1-4\pi C^\rA(r_{12}=2)\left/\int\rd^2\br_2\,C^\rA(\br_{12})\right.\nonumber\\
    \tilde{\Delta}_{\phi,\textrm{spec.}}&=3\frac{E_\phi-E_0}{E_{T^{\mu\nu}}-E_0}.
    \label{eq:dim}
\end{align}
At a conformal region, these quantities are size-independent and have the same value as the scaling dimension $\Delta_\phi$; towards the symmetry-broken phase, these quantities decrease with $\Norb$ towards $0$; towards the disordered phase, these quantities increase with system size. Numerically, we add a perturbation to $U_2$ and study their dependence on $\delta U_2$~(Fig.~\ref{fig:fss}). We indeed find that these quantities have a crossing around $\delta U_2=0$ and the slope is increasing with system size, which suggests that this CFT describes a phase transition controlled by the slightly relevant operator $S$: $\delta U_2>0$ drives it towards a symmetry-broken phase and $\delta_2<0$ drives it towards a disordered phase. However, we need to note that these behaviours are not entirely consistent: The $\tilde{\Delta}_{S,\textrm{spec.}}=3(E_S-E_0)/(E_{T^{\mu\nu}}-E_0)$ show non-monotonic dependence at different $\Norb$. This inconsistency can be reconciled by taking into account the contributions of the irrelevant operators.

\begin{figure}[htbp]
    \centering
    \includegraphics[width=\figurewidth]{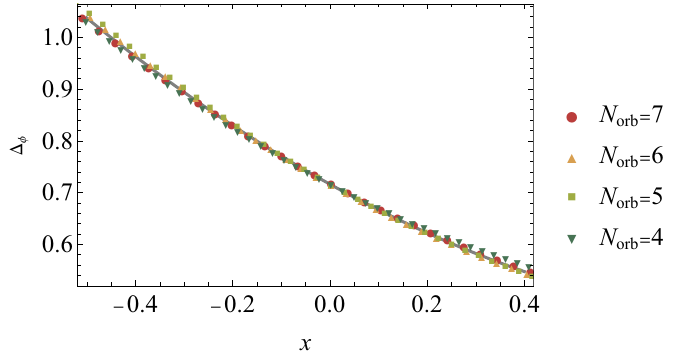}
    \caption{The data collapse of $\tilde{\Delta}_{\phi,\textrm{spec}}$ plotted against $x=(U_2-U_{2,0})\Norb^{(3-\Delta_S)/2}$ for different system sizes.}
    \label{fig:colps}
\end{figure}

We perform a data collapse for Fig.~\ref{fig:fss}c under the finite-size scaling hypothesis that 
\begin{equation}
    \tilde{\Delta}_{\phi,\textrm{spec}}(U_2)=f((U_2-U_{2,0})\Norb^{(3-\Delta_S)/2}).
\end{equation}
Specifically, we take the ansatz 
\begin{align}
    x&=(\delta U_2-\delta U_{2,0})\Norb^{(3-\Delta_S)/2}\nonumber\\
    \tilde{\Delta}_{\phi,\textrm{spec}}&=C_0+C_1x+C_2x^2+C_3x^3
\end{align}
and perform a fitting with respect to $\delta U_{2,0},\Delta_S,C_{0,1,2,3}$. The result is 
\begin{equation*}
    \delta U_{2,0}=0.0292,\ \Delta_S=1.7386,\ C_0=0.716,\ C_1=-0.519,\ C_2=0.274,\ C_3=-0.005.
\end{equation*}
The $\tilde{\Delta}_{\phi,\textrm{spec}}$ plotted against $x$ for different system sizes exhibit good collapse (Fig.~\ref{fig:colps}). However, the fitted $\Delta_S$ is inconsistent with the value $\Delta_S=2.98$ by state-operator correspondence. These inconsistencies indicate that the finite-size behaviours are subject to large contributions from the irrelevant operators, and a more careful analysis is needed to determine the stability of the CFT and the property of the proximate phases. 

\FloatBarrier

\section{Parameter dependence of the scaling dimensions}
\label{app:dim}

\begin{figure}[htbp]
    \centering
    \includegraphics[width=\linewidth]{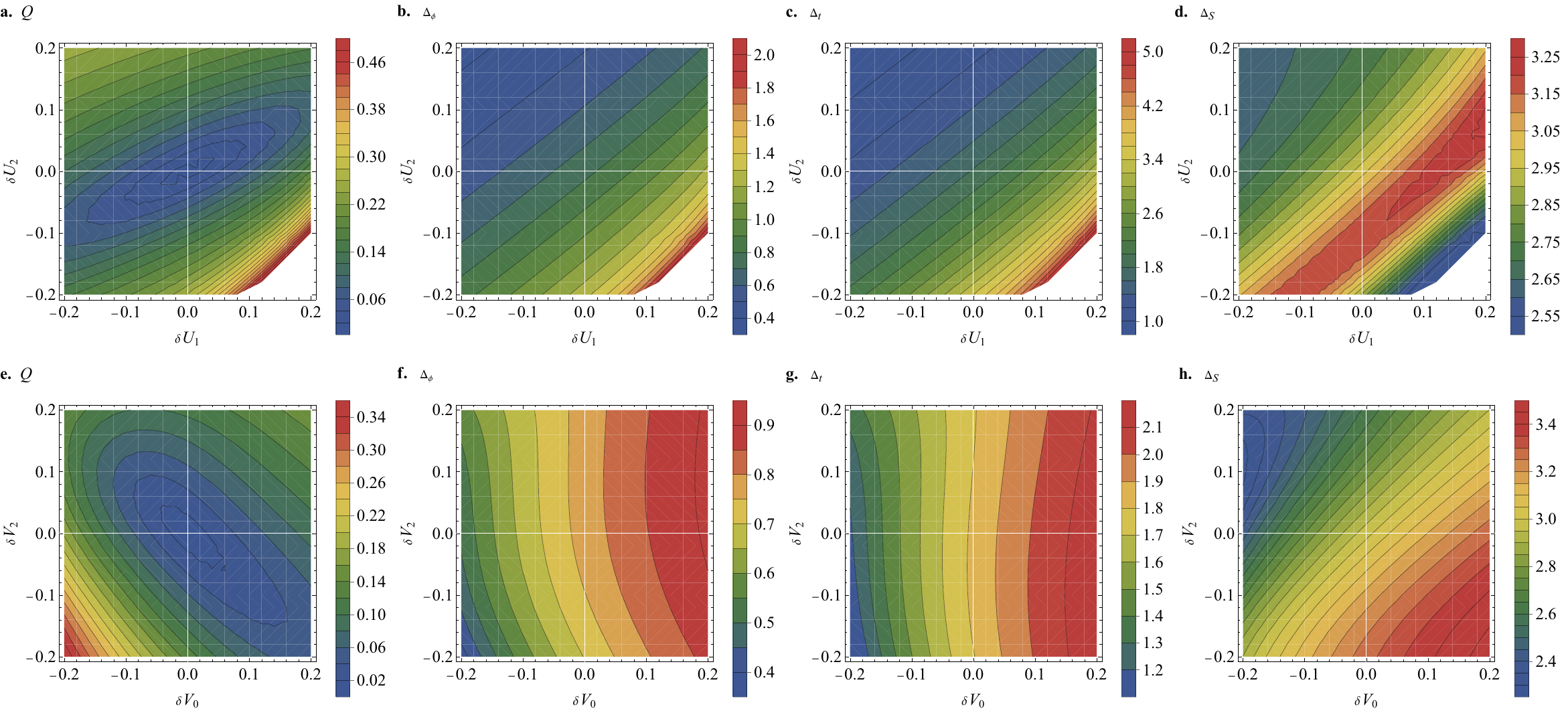}
    \caption{The parameter dependence of cost function $Q$ and scaling dimensions of $\phi_{(1,1)}$, $t_{(2,2)}$ and $S$ on the $U_1$-$U_2$ plane and $V_0$-$V_2$ plane for the $(N,M)=(3,1)$ model and $\Norb=7$.}
    \label{fig:lsc}
\end{figure}

\begin{figure}[htbp]
    \centering
    \includegraphics[width=\linewidth]{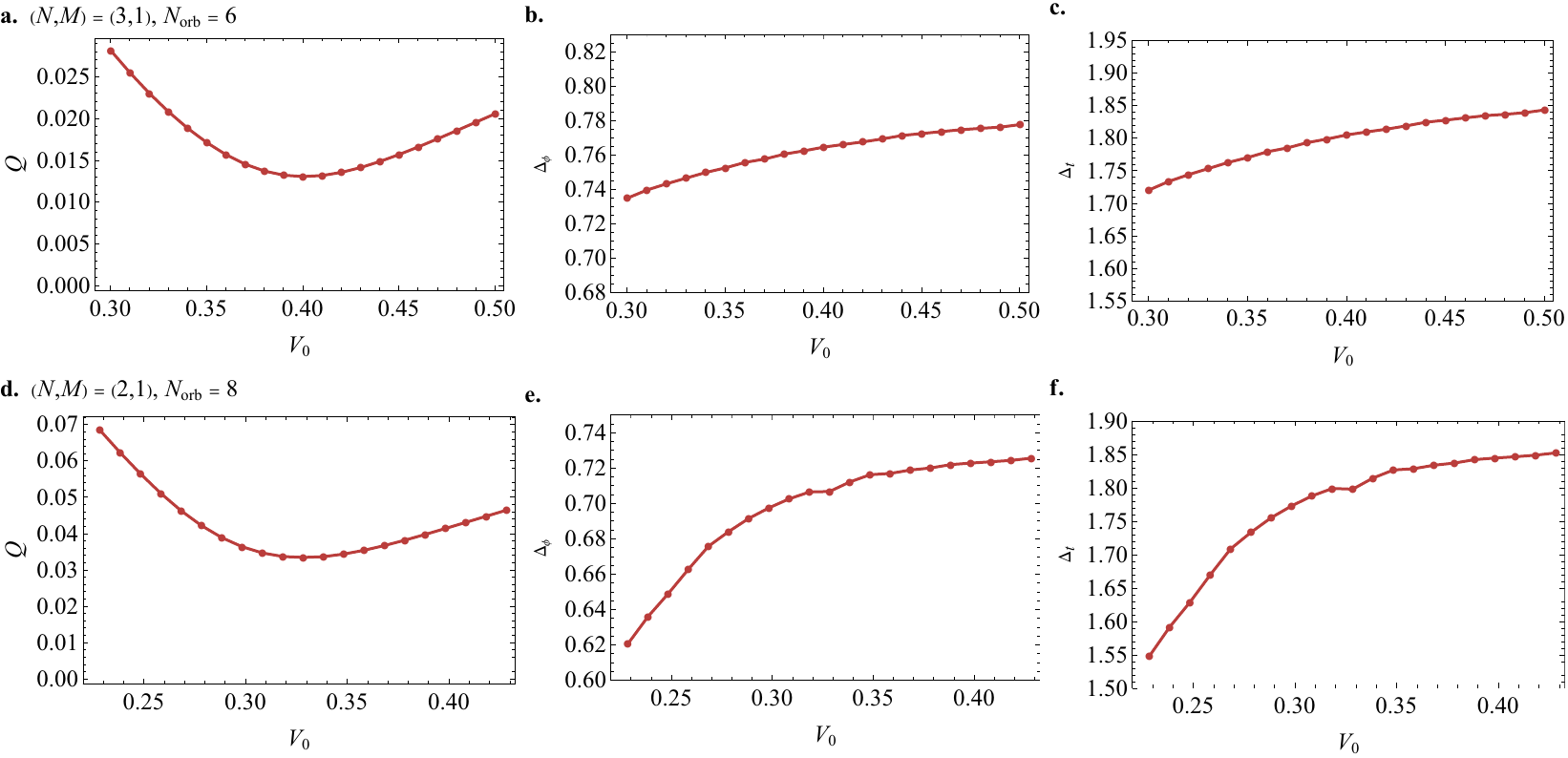}
    \caption{The parameter dependence of cost function $Q$ and scaling dimensions of $\phi_{(1,1)}$, $t_{(2,2)}$ in the $(N,M)=(3,1)$ (upper line) and $(N,M)=(2,1)$ (lower line) theories. The data are obtained fixing the parameter $V_0$ at different values and optimise over the rest of the parameters $U_{1,2}$ and $V_2$. The scale of coordinates in b and e, c and f are kept to be the same to facilitate comparison. }
    \label{fig:lsc_opt}
\end{figure}

\FloatBarrier

\section{Full spectrum}
\label{app:full_spec}

\begin{longtable}[htbp]{ccc|ccc|cccc}
    \caption{The scaling dimensions of the first 100 states for the theories $(N,M)=(3,1),(4,1),(4,2)$}\label{tbl:full_spec}\\[3pt]
    \hline\hline
    \vphantom{x}&&&&&&&&&\\
    \multicolumn{3}{c|}{$(N,M)=(3,1)$}&\multicolumn{3}{c|}{$(N,M)=(4,1)$}&\multicolumn{4}{c}{$(N,M)=(4,2)$}\\[1pt]
    $\Delta$&$L^2$&$C_2$&$\Delta$&$L^2$&$C_2$&$\Delta$&$L^2$&$C_2$&$\mathcal{P}$\\[1pt]
    \hline
    \vphantom{x}&&&&&&&&&\\
    \endfirsthead
    \caption{(Continued)}\\[3pt]
    \hline\hline
    \vphantom{x}&&&&&&&&&\\
    \multicolumn{3}{c|}{$(N,M)=(3,1)$}&\multicolumn{3}{c|}{$(N,M)=(4,1)$}&\multicolumn{4}{c}{$(N,M)=(4,2)$}\\[1pt]
    $\Delta$&$L^2$&$C_2$&$\Delta$&$L^2$&$C_2$&$\Delta$&$L^2$&$C_2$&$\mathcal{P}$\\[1pt]
    \hline
    \vphantom{x}&&&&&&&&&\\
    \endhead
    \hline\hline
    \endfoot
    0.0000 & 0 & 0 & 0.0000 & 0 & 0 & 0.0000 & 0 & 0 & $+$ \\[3pt]
    0.7695 & 0 & 3 & 0.7951 & 0 & 4 & 0.7824 & 0 & 4 & $-$ \\[3pt]
    1.7867 & 2 & 3 & 1.7978 & 0 & 9 & 1.1719 & 0 & 6 & $+$ \\[3pt]
    1.8182 & 0 & 7 & 1.8060 & 2 & 4 & 1.7672 & 0 & 9 & $+$ \\[3pt]
    2.0049 & 2 & 4 & 2.0082 & 2 & 5 & 1.7902 & 2 & 4 & $-$ \\[3pt]
    2.5213 & 2 & 6 & 2.5856 & 2 & 8 & 2.0025 & 2 & 5 & $+$ \\[3pt]
    2.7692 & 6 & 3 & 2.7747 & 2 & 9 & 2.1600 & 0 & 11 & $-$ \\[3pt]
    2.7754 & 0 & 3 & 2.7973 & 0 & 4 & 2.1915 & 2 & 6 & $+$ \\[3pt]
    2.7998 & 2 & 7 & 2.8030 & 6 & 4 & 2.5544 & 2 & 8 & $+$ \\[3pt]
    2.9787 & 0 & 0 & 2.9586 & 0 & 15 & 2.5625 & 2 & 8 & $-$ \\[3pt]
    2.9907 & 2 & 4 & 2.9874 & 0 & 0 & 2.7334 & 2 & 9 & $+$ \\[3pt]
    2.9976 & 6 & 0 & 2.9945 & 6 & 5 & 2.7569 & 0 & 14 & $+$ \\[3pt]
    3.0049 & 6 & 4 & 2.9973 & 2 & 5 & 2.7922 & 0 & 4 & $-$ \\[3pt]
    3.0786 & 0 & 12 & 2.9998 & 6 & 0 & 2.7989 & 6 & 4 & $-$ \\[3pt]
    3.2624 & 2 & 4 & 3.2042 & 2 & 10 & 2.9157 & 0 & 15 & $-$ \\[3pt]
    3.2686 & 2 & 8 & 3.2891 & 2 & 5 & 2.9609 & 0 & 6 & $+$ \\[3pt]
    3.3127 & 0 & 3 & 3.3360 & 6 & 4 & 2.9919 & 6 & 0 & $+$ \\[3pt]
    3.3417 & 6 & 3 & 3.3900 & 0 & 4 & 2.9936 & 6 & 5 & $+$ \\[3pt]
    3.4185 & 2 & 6 & 3.4078 & 0 & 6 & 3.0034 & 2 & 5 & $-$ \\[3pt]
    3.4187 & 0 & 6 & 3.4747 & 12 & 4 & 3.0282 & 0 & 0 & $+$ \\[3pt]
    3.5330 & 6 & 6 & 3.4772 & 2 & 8 & 3.1143 & 2 & 11 & $-$ \\[3pt]
    3.5418 & 2 & 6 & 3.4872 & 0 & 8 & 3.1543 & 2 & 11 & $-$ \\[3pt]
    3.5871 & 12 & 3 & 3.5691 & 6 & 9 & 3.1648 & 2 & 10 & $-$ \\[3pt]
    3.6451 & 6 & 7 & 3.5968 & 2 & 8 & 3.1811 & 0 & 4 & $+$ \\[3pt]
    3.6737 & 6 & 7 & 3.6073 & 6 & 8 & 3.2081 & 6 & 6 & $+$ \\[3pt]
    3.7185 & 6 & 8 & 3.6591 & 12 & 5 & 3.2384 & 2 & 5 & $+$ \\[3pt]
    3.7355 & 2 & 3 & 3.6852 & 6 & 10 & 3.2857 & 0 & 8 & $-$ \\[3pt]
    3.7471 & 2 & 10.5 & 3.6869 & 6 & 9 & 3.3024 & 0 & 17 & $+$ \\[3pt]
    3.8318 & 0 & 7 & 3.6968 & 6 & 6 & 3.3291 & 2 & 12 & $+$ \\[3pt]
    3.8438 & 12 & 4 & 3.7251 & 2 & 13.5 & 3.3684 & 6 & 4 & $+$ \\[3pt]
    3.8613 & 6 & 4 & 3.7775 & 20 & 4 & 3.3826 & 2 & 8 & $-$ \\[3pt]
    3.8971 & 12 & 0 & 3.8291 & 6 & 5 & 3.4202 & 12 & 4 & $-$ \\[3pt]
    3.9686 & 6 & 3 & 3.8483 & 2 & 4 & 3.4205 & 2 & 8 & $+$ \\[3pt]
    3.9735 & 2 & 4 & 3.8750 & 30 & 4 & 3.4902 & 0 & 8 & $+$ \\[3pt]
    3.9797 & 0 & 7 & 3.8787 & 2 & 15 & 3.5024 & 6 & 9 & $+$ \\[3pt]
    4.0029 & 2 & 12 & 3.8865 & 12 & 0 & 3.5287 & 0 & 4 & $-$ \\[3pt]
    4.0074 & 6 & 0 & 3.9339 & 0 & 9 & 3.5349 & 6 & 4 & $-$ \\[3pt]
    4.0207 & 2 & 0 & 3.9440 & 20 & 5 & 3.5396 & 0 & 6 & $+$ \\[3pt]
    4.0929 & 2 & 8 & 4.0180 & 30 & 5 & 3.579 & 0 & 9 & $+$ \\[3pt]
    4.1087 & 6 & 3 & 4.0199 & 6 & 4 & 3.5804 & 6 & 8 & $-$ \\[3pt]
    4.1235 & 2 & 9 & 4.0796 & 6 & 0 & 3.5827 & 6 & 8 & $+$ \\[3pt]
    4.1854 & 6 & 8 & 4.0837 & 6 & 10 & 3.5891 & 12 & 5 & $+$ \\[3pt]
    4.1872 & 6 & 6 & 4.0859 & 2 & 5 & 3.6017 & 2 & 12 & $+$ \\[3pt]
    4.1985 & 20 & 3 & 4.0911 & 0 & 9 & 3.603 & 2 & 8 & $+$ \\[3pt]
    4.2085 & 6 & 4 & 4.1208 & 6 & 4 & 3.6149 & 20 & 4 & $-$ \\[3pt]
    4.2143 & 2 & 3 & 4.1223 & 2 & 10 & 3.6159 & 6 & 6 & $+$ \\[3pt]
    4.2210 & 2 & 8 & 4.1441 & 0 & 11 & 3.6407 & 2 & 13.5 & $-$ \\[3pt]
    4.2327 & 2 & 6 & 4.1480 & 12 & 9 & 3.6824 & 6 & 10 & $-$ \\[3pt]
    4.2378 & 0 & 4 & 4.1858 & 2 & 10 & 3.7004 & 6 & 9 & $+$ \\[3pt]
    4.2962 & 2 & 4 & 4.2040 & 6 & 8 & 3.7234 & 2 & 13.5 & $+$ \\[3pt]
    4.3289 & 2 & 3 & 4.2090 & 2 & 0 & 3.7297 & 2 & 14 & $+$ \\[3pt]
    4.33 & 6 & 3 & 4.2254 & 2 & 12 & 3.7356 & 2 & 8 & $-$ \\[3pt]
    4.3484 & 0 & 8 & 4.2391 & 6 & 5 & 3.7666 & 20 & 5 & $+$ \\[3pt]
    4.3557 & 2 & 6 & 4.2509 & 0 & 22 & 3.7919 & 0 & 8 & $-$ \\[3pt]
    4.3714 & 12 & 7 & 4.2856 & 12 & 8 & 3.8001 & 12 & 6 & $+$ \\[3pt]
    4.3757 & 12 & 4 & 4.2964 & 2 & 8 & 3.8096 & 6 & 5 & $-$ \\[3pt]
    4.3764 & 12 & 6 & 4.2992 & 2 & 12 & 3.8171 & 2 & 15 & $-$ \\[3pt]
    4.3859 & 12 & 3 & 4.3007 & 2 & 4 & 3.8329 & 2 & 4 & $-$ \\[3pt]
    4.3860 & 6 & 6 & 4.3128 & 12 & 5 & 3.8826 & 0 & 20 & $-$ \\[3pt]
    4.4199 & 6 & 7 & 4.3265 & 12 & 5 & 3.8834 & 0 & 11 & $-$ \\[3pt]
    4.4208 & 0 & 6 & 4.3455 & 0 & 5 & 3.8921 & 6 & 11 & $-$ \\[3pt]
    4.4424 & 6 & 10.5 & 4.3682 & 6 & 8 & 3.8955 & 0 & 9 & $-$ \\[3pt]
    4.4495 & 20 & 4 & 4.3849 & 6 & 9 & 3.8964 & 2 & 10 & $+$ \\[3pt]
    4.4527 & 6 & 6 & 4.3859 & 12 & 4 & 3.944 & 2 & 11 & $+$ \\[3pt]
    4.4826 & 6 & 6 & 4.4078 & 6 & 4 & 3.9528 & 0 & 9 & $+$ \\[3pt]
    4.5126 & 0 & 18 & 4.4195 & 6 & 13.5 & 3.9752 & 6 & 8 & $-$ \\[3pt]
    4.5263 & 0 & 7 & 4.4379 & 20 & 9 & 3.9852 & 2 & 11 & $-$ \\[3pt]
    4.5477 & 2 & 6 & 4.4437 & 2 & 6 & 3.9865 & 12 & 0 & $+$ \\[3pt]
    4.5505 & 12 & 4 & 4.4454 & 12 & 10 & 3.9979 & 0 & 12 & $+$ \\[3pt]
    4.5658 & 0 & 3 & 4.4533 & 2 & 5 & 4.0064 & 20 & 6 & $+$ \\[3pt]
    4.5664 & 6 & 3 & 4.4627 & 12 & 9 & 4.0073 & 2 & 10 & $-$ \\[3pt]
    4.5779 & 30 & 3 & 4.4667 & 6 & 8 & 4.0076 & 12 & 9 & $+$ \\[3pt]
    4.5807 & 0 & 0 & 4.5051 & 0 & 10 & 4.0105 & 2 & 6 & $+$ \\[3pt]
    4.5818 & 12 & 7 & 4.5171 & 2 & 4 & 4.0163 & 6 & 4 & $-$ \\[3pt]
    4.5981 & 2 & 10.5 & 4.5191 & 0 & 8 & 4.0233 & 6 & 10 & $-$ \\[3pt]
    4.6157 & 2 & 7 & 4.5199 & 2 & 8 & 4.0344 & 6 & 8 & $-$ \\[3pt]
    4.630 & 2 & 4 & 4.5242 & 2 & 16 & 4.046 & 6 & 4 & $+$ \\[3pt]
    4.6384 & 12 & 8 & 4.5466 & 6 & 8 & 4.0515 & 0 & 12 & $+$ \\[3pt]
    4.6504 & 20 & 0 & 4.5478 & 20 & 5 & 4.0562 & 6 & 12 & $+$ \\[3pt]
    4.6551 & 6 & 7 & 4.5604 & 30 & 0 & 4.0791 & 2 & 5 & $+$ \\[3pt]
    4.6673 & 12 & 6 & 4.5673 & 20 & 8 & 4.0812 & 6 & 11 & $-$ \\[3pt]
    4.6836 & 6 & 8 & 4.5735 & 2 & 13.5 & 4.0907 & 2 & 12 & $+$ \\[3pt]
    4.6889 & 0 & 10.5 & 4.5789 & 2 & 6 & 4.1010 & 6 & 5 & $+$ \\[3pt]
    4.6891 & 12 & 4 & 4.592 & 6 & 15 & 4.1061 & 6 & 0 & $-$ \\[3pt]
    4.6905 & 6 & 10.5 & 4.5957 & 0 & 9 & 4.1115 & 2 & 16 & $+$ \\[3pt]
    4.7003 & 12 & 0 & 4.5996 & 12 & 8 & 4.1147 & 2 & 12 & $-$ \\[3pt]
    4.7004 & 2 & 8 & 4.6299 & 6 & 10 & 4.1209 & 6 & 8 & $+$ \\[3pt]
    4.7081 & 2 & 13 & 4.6412 & 12 & 10 & 4.1242 & 2 & 4 & $-$ \\[3pt]
    4.7248 & 2 & 8 & 4.6415 & 30 & 8 & 4.1301 & 2 & 10 & $+$ \\[3pt]
    4.7321 & 2 & 4 & 4.6482 & 2 & 9 & 4.1366 & 2 & 8 & $-$ \\[3pt]
    4.7369 & 6 & 0 & 4.6535 & 2 & 10 & 4.1368 & 6 & 9 & $+$ \\[3pt]
    4.7420 & 42 & 3 & 4.6786 & 6 & 13.5 & 4.1775 & 0 & 11 & $+$ \\[3pt]
    4.7429 & 2 & 7 & 4.6921 & 6 & 9 & 4.1814 & 12 & 8 & $-$ \\[3pt]
    4.7585 & 2 & 6 & 4.6933 & 6 & 6 & 4.1968 & 2 & 17 & $+$ \\[3pt]
    4.7606 & 6 & 4 & 4.7069 & 2 & 8 & 4.1991 & 2 & 4 & $+$ \\[3pt]
    4.7751 & 6 & 12 & 4.7129 & 12 & 8 & 4.2027 & 0 & 22 & $+$ \\[3pt]
    4.7793 & 0 & 10 & 4.7192 & 0 & 13.5 & 4.2029 & 12 & 8 & $+$ \\[3pt]
    4.7797 & 2 & 3 & 4.7265 & 6 & 5 & 4.2098 & 6 & 11 & $-$ \\[3pt]
    4.7822 & 6 & 4 & 4.7343 & 6 & 11 & 4.2158 & 12 & 5 & $-$ \\[3pt]
    4.7923 & 12& 6 & 4.7642 & 12 & 4 & 4.2193 & 2 & 0 & $+$ \\[3pt]
\end{longtable}

\end{document}